\documentclass[11pt]{article}
\usepackage{amsmath,amsthm,amssymb,euscript,epsf,epsfig}
\usepackage{array}
\makeatletter
\@addtoreset{equation}{section}
\makeatother

\def\one{{\hbox{ 1\kern-.8mm l}}}
\def\zero{{\hbox{ 0\kern-1.5mm 0}}}

\def\hs{\hat{s}}

\def\tr{\tilde R}

\newcommand{\sm}[1]{\mbox{\scriptsize #1}}
\newcommand{\tn}[1]{\mbox{\tiny #1}}
\renewcommand{\@}[1]{\sqrt{#1}}
\newcommand{\Tr}{{\mbox{Tr}}}
\renewcommand{\tr}{{\mbox{tr}}}
\renewcommand{\le}[1]{\label{#1}\end{eqnarray}}
\newcommand{\be}{\begin{equation}}
\newcommand{\ee}{\end{equation}}
\newcommand{\bea}{\begin{eqnarray}}
\newcommand{\eea}{\end{eqnarray}}
\newcommand{\nn}{\nonumber\\}

\newcommand{\eq}[1]{(\ref{#1})}
\def\nn{\nonumber\\}
\def\nm{\nonumber}

\def\ffract#1#2{\raise .35 em\hbox{$\scriptstyle#1$}\kern-.25em/
\kern-.2em\lower .22 em \hbox{$\scriptstyle#2$}}

\def\half{{1\over2}\,}
\def\dd{\mbox{d}}

\def\O{\Omega}

\def\bra{\langle}
\def\ket{\rangle}
\def\eht{\hat{t}}

\def\d{\delta}

\def\g{\gamma}

\def\l{\lambda}

\def\s{\sigma}
\def\S{\Sigma}

\def\r{\rho}

\def\cR{ { \check R } }
\newdimen\tableauside\tableauside=1.0ex
\newdimen\tableaurule\tableaurule=0.4pt
\newdimen\tableaustep
\def\phantomhrule#1{\hbox{\vbox to0pt{\hrule height\tableaurule width#1\vss}}}
\def\phantomvrule#1{\vbox{\hbox to0pt{\vrule width\tableaurule height#1\hss}}}
\def\sqr{\vbox{%
  \phantomhrule\tableaustep
  \hbox{\phantomvrule\tableaustep\kern\tableaustep\phantomvrule\tableaustep}%
  \hbox{\vbox{\phantomhrule\tableauside}\kern-\tableaurule}}}
\def\squares#1{\hbox{\count0=#1\noindent\loop\sqr
  \advance\count0 by-1 \ifnum\count0>0\repeat}}
\def\tableau#1{\vcenter{\offinterlineskip
  \tableaustep=\tableauside\advance\tableaustep by-\tableaurule
  \kern\normallineskip\hbox
    {\kern\normallineskip\vbox
      {\gettableau#1 0 }%
     \kern\normallineskip\kern\tableaurule}%
  \kern\normallineskip\kern\tableaurule}}
\def\gettableau#1 {\ifnum#1=0\let\next=\null\else
  \squares{#1}\let\next=\gettableau\fi\next}

\begin{document}

%{}~
%{}~

\rightline{AEI-2005-162}
\rightline{HU-EP-06/09}
\rightline{QMUL-PH-05-16}

\vskip2cm

\centerline{{\Large \bf Large $N$ Expansion of   $q$-Deformed
  Two-Dimensional}} 
\vskip .5cm
\centerline{{ \bf \Large  Yang-Mills Theory  and Hecke Algebras   }}

\medskip

\vspace*{4.0ex}
\vskip 1cm

\centerline{\large \bf
Sebastian de Haro${}^1$, Sanjaye Ramgoolam${}^2$, Alessandro Torrielli${}^3$ }

\vspace*{4.0ex}
\begin{center}
{\large ${}^1$ Max-Planck-Institut f\"ur Gravitationsphysik \\
Albert-Einstein-Institut\\
14476 Golm, Germany \\}
\vskip.1in 
{\large ${}^2$Department of Physics\\
Queen Mary, University of London\\
Mile End Road,
London E1 4NS UK\\}
\vskip.1in
{\large ${}^3$  Institut f\"ur Physik, Humboldt Universit\"at zu Berlin \\
Newtonstra\ss e 15, D-12489 Berlin, Germany}
\vskip.5cm
%\vfill
%\begin{flushright}
{\tt {sdh@aei.mpg.de,  s.ramgoolam@qmul.ac.uk,  torriell@physik.hu-berlin.de }} 
%\end{flushright}
\end{center}

\vspace*{5.0ex}
\vskip2cm

\centerline{\bf Abstract} \bigskip

 We derive the $q$-deformation of the chiral Gross-Taylor 
 holomorphic string large 
$N$  expansion of two dimensional $SU(N)$ Yang-Mills theory. 
 Delta functions on  symmetric group algebras
 are replaced by the corresponding objects (canonical trace functions) 
 for Hecke algebras. 
 The role of the Schur-Weyl duality   
 between unitary groups and symmetric groups is now played by 
 $q$-deformed Schur-Weyl duality of quantum groups. 
 The appearance of Euler characters of configuration spaces 
 of Riemann surfaces in the expansion persists. We discuss 
 the geometrical meaning of these formulae.

\eject

\tableofcontents

\newpage

\section{Introduction and summary of the results}

Two-dimensional Yang-Mills theory, on a Riemann surface 
of genus $G$ and of area $A$,  can be solved exactly. The partition function 
is 
\be\label{2dYM}
Z_{\sm{YM}}(G,A )  = \sum_R   ( \mbox{dim}(R))^{2-2G } e^{-g_{\tn{YM}}^2  A\, C_2 ( R )  }~.
\ee  
This result was first obtained using the lattice formulation, 
followed by a continuum limit \cite{migdal}.  The sum is over all 
irreducible representations of the gauge group, the cases $U(N)$ or 
$SU(N)$ will be of interest here.

Gross and Taylor  \cite{gross,gti,gtii} studied the large $N$ expansion 
of two-dimensional Yang-Mills theory with gauge group 
$U(N)$ and $SU(N)$ and showed that it is equivalent to a
string theory.  They showed that the large $N$ 
expansion is given by a non-chiral expansion, which is 
a sum involving chiral and anti-chiral factors.
The chiral expansion of  \eq{2dYM}\footnote{In this paper we work at zero area. The computations can
be generalized to the case of finite area along the lines of \cite{gti,gtii}.}
is given by 
\be\label{cgt}
Z_{\sm{YM}}^+(G)=\sum_{n=0}^\infty {1\over N^{(2G-2)n}}\sum_{s_i,t_i\in S_n}
{1\over n!}\,\d\left(\O^{2-2G}_n
\prod_{i=1}^Gs_it_is_i^{-1}t_i^{-1}\right)~.
\ee
It is a sum consisting of  delta functions over symmetric groups, 
which count homomorphisms from the fundamental group 
of punctured Riemann surfaces to the symmetric groups. These 
homomorphisms are known to count  branched covers of $ \Sigma_G $.  
It was shown in \cite{cmri,cmrii} that the chiral sum actually computes
an Euler character of moduli spaces of holomorphic maps
with fixed target space. This was done by expanding the 
$\Omega $ factors, and recognising that the coefficients in the 
expansion are Euler characters of configuration spaces 
of (branch) points on $ \Sigma_G$. Topological string theory 
constructions were then used to derive a path integral which 
localizes to an integral of the Euler class 
on the moduli space of holomorphic maps. 
 For simplicity we are discussing only the chiral part of the
 partition function here, but there is an analogous expansion for the
full partition function. 
A different string action involving harmonic maps was proposed 
in \cite{horava}.

 Two-dimensional Yang-Mills has recently found a surprising
 new application in connection with topological strings on a non-compact 
Calabi-Yau and black hole entropy \cite{vafa}.
 The $q$-deformation of two-dimensional Yang-Mills has also
 found an application in this context \cite{ bp,aosv}.
 The partition function of $q$-deformed 
Yang-Mills has been obtained by replacing the scalar dual field 
of the Yang-Mills field strength by a compact scalar. 
Such a compact scalar is natural from the point of view of the
 worldvolume of D4-branes wrapping a 4-cycle of the non-compact Calabi-Yau. 
 New connections with Turaev invariants have 
 also been suggested  \cite{Sebaslps}. The $q$-deformation of
 two-dimensional Yang-Mills theory was studied earlier
\cite{boulatov,br} (see also \cite{klimcik}).   

The $q$-deformation of the zero area partition function of
two-dimensional Yang-Mills  is 
\bea\label{qpartf}
Z_{\sm{$q$YM}} ( G )   = \sum_R   ( \mbox{dim}_q\, R )^{2-2G }~.
\eea
In the context of \cite{aosv} this is the limit where the degree $p$  
of one of the line bundles  is  zero. 
In the $q$-deformed Yang-Mills, the universal enveloping 
algebra of $U(N)$ is replaced by $U_q ( u(N) )$. 
The exact partition function for a closed Riemann surface, 
which is expressed in terms of dimensions of irreducible representations
of $U(N)$, is now expressed in terms of $q$-dimensions
of $U_q ( u(N) )$ representations. The same remarks apply to 
$U_q ( su(N) )$.  

The underlying algebraic relation  which leads to the 
relation between the sum over $U(N)$ representations in (\ref{2dYM})
and the delta functions over symmetric groups in (\ref{cgt})
is Schur-Weyl duality, which we describe further in section 2. 
The $q$-deformation of the Schur-Weyl duality  between 
$U(N)$ and $S_n$ is known \cite{jimbo}. In this $q$-deformation, 
the role of the group algebra of $S_n$ ( denoted by $ \mathbb{C} S_n $ ) 
 is played by the Hecke algebra $H_n(q)$. 

In this paper, we show that the large $N$, chiral 
Gross-Taylor expansion,  in terms of symmetric group data 
can be $q$-deformed to give an expansion in terms of Hecke algebra data. 
In this case we find the following result:
\be\label{expansion}
Z_{q\sm{YM}}(G)=\sum_{n=0}^\infty \sum_{s_it_i \in S_n }\,{1\over g}\,[N]^{(2-2G)n}\,
\d\left(D\,\Omega_n^{2-2G}\prod_{i=1}^Gq^{-l(s_i)-l(t_i)}
h(s_i)h(t_i)h(s_i^{-1})h(t_i^{-1})\right)~ 
\ee
Here, $h(s)\in H_n$ is the Hecke algebra element associated to $s\in S_n$. 
That such an expansion is possible at all in the quantum
case is highly non-trivial and very much suggestive of a geometric
 interpretation in terms of deformations of maps, on which we comment
in section \ref{discussion}. The possibility of the expansion \eq{expansion} 
depends crucially on the existence of suitable central elements of the
Hecke algebra (like $D$ and $\O_n$, to be defined later). These central
 elements play an important role in that they also determine the data
on manifolds with closed boundary:
\bea
Z(\S_G;C_1,\ldots,C_B)&=&\sum_R\,[N]^{(2-2G-B)n}\sum_{s_it_i}{1\over g}\,
\times\nn
\times\!\!\!&\d&\!\!\!\!\left(D^{1-B}\,\O_n^{2-2G-B}
\prod_{i=1}^Gq^{-l(s_i)-l(t_i)}h(s_i)h(t_i)h(s_i^{-1})h(t_i^{-1})
\prod_{j=1}^BC_j\right)~.
\eea
In this formula, the central elements of the Hecke algebra take
 over the role of the holonomies of the gauge field around the
$B$ boundaries of $\S_G$. 

We also work out the case of non-intersecting Wilson loops.
 We develop an analog of the Verlinde formula for the tensor product
multiplicity coefficients of $SU(N)$ in terms of characters
 of the Hecke algebra. To our knowledge, this formula has not appeared in 
the literature. Expectation values of Wilson loops can now 
again be written as Hecke delta functions which are natural deformations of the
symmetric group delta functions.

In four appendices we give some of the facts and proofs about
 Hecke algebras that we use in the main text. To our knowledge,
 some of the 
formulas proven in these appendices are not available in the
 mathematical literature before.

\section{Hecke algebras and the chiral expansion of $q$-deformed 2dYM} 

\subsection{Review of the Gross-Taylor expansion}

Before we do the $q$-deformed case, we will review the main tools used in the derivation of the partition function of 2d Yang-Mills as a 
topological theory counting branched covers of the Riemann surface. For full details we refer to \cite{cmri}. For simplicity, we discuss the 
case of zero-area and no Wilson loops in this section. We start writing out the partition function as a sum over Young tableaux:
\be
Z_{\sm{2dYM}}(\S_G;A)=\sum_R\,(\mbox{dim}(R))^{2-2G}=\sum_{n=0}^\infty\sum_{Y\in {\cal Y}_n^N}\left(\mbox{dim}(R(Y))\right)^{2-2G}~,
\ee
where we sum over the set ${\cal Y}_n^N$ of $SU(N)$ Young diagrams with $n$ boxes and number of rows less than $N$. Of course, we also sum over diagrams with arbitrary number of boxes. The chiral expansion is derived by dropping the constraint on the number of rows. Next we use Schur-Weyl duality to derive the following fomula:
\be
\mbox{dim}(R)={N^n\over n!}\,\chi_R(\Omega_n)~.
\ee
We are using a notation where $R=R(Y)$ denotes both the $SU(N)$ and
 the $S_n$ representation corresponding to a Young tableau with $n$ boxes,
 $Y$. $\chi_R$ is a character of the symmetric group, and $\O_n$ is a 
particular central element in $ \mathbb { C } S_n$ given in \cite{gti,gtii}.
The chiral Gross-Taylor expansion is obtained as  
\bea\label{gtexp} 
Z_{\sm{2dYM}}(\S_G;A)&=&\sum_{n=0}^\infty\sum_R N^{(2-2G)n}
\left(d_R\over n!\right)^{2-2G}{1\over d_R}\,\chi_R(\O^{2-2G}_n)\nn
&=&
\sum_{n=0}^\infty N^{(2-2G)n}{1\over n!}\sum_{s_i,t_i\in S_n}\d
\left(\O^{2-2G}_n\prod_{i=1}^Gs_it_is_i^{-1}t_i^{-1} \right)~.
\eea
The fact that  $ \Omega_n  $ is a central element  in the group 
algebra $ \mathbb { C } S_n $ is important. 
This is explained in more detail and  generalized to
the $q$-deformed case in section (2.3). 
Another important identity which enters \eq{gtexp} 
is 
\be 
\sum_{s,t \in S_n } { 1\over d_R }\,\chi_R  ( s t s^{-1} t^{-1} ) = 
 \left({n! \over d_R }\right)^2 ~,
\ee 
where it is easy to see that $ \sum_{s,t}  s t s^{-1} t^{-1} $ 
is a central element of  $ \mathbb { C } S_n $. We find 
\eq{qdefpirel}, which gives the $q$-deformation of this equation, 
and we prove related centrality properties for $H_n(q) $ in Appendix A.

\subsection{Hecke algebras and Schur-Weyl duality}

There is a natural generalization of the previous formulas 
using Hecke algebras. In this subsection we review basic facts
 about Hecke algebras and derive some formulas that
 we will use in what follows.

The symmetric group $S_n$ can be defined in terms of generators $s_i$ ($i = 1,\ldots, n-1 $), which obey relations 
\bea\label{gensym} 
s_i^2 &=& 1 \,\, ~~~~~~~~~~~~~~~~\hbox{for} ~~ i = 1,\ldots, n-1 \nn 
s_i\, s_{i+1}\, s_i &=&   s_{i+1}\, s_i\, s_{i+1}\, ~~~~ \hbox{for} ~~  i = 1,\ldots, n-2 \nn 
s_i\, s_j &=& s_j\, s_i\,\, ~~~~~~~~~~~~ \hbox{for} ~~  | i - j | \ge 2 
\eea
The minimal length of a word in the $s_i$ which is equal to 
a permutation $\sigma$ is called the length of the permutation
and is denoted as $l(\sigma)$.

The Hecke algebra $H_n(q)$  is defined in terms of generators $g_i$ 
which obey \cite{kingwyb}
\bea\label{hecke}
g_i^2 &=& (q-1)\, g_i + q\,\,\,  ~~ \hbox{for} ~~ i = 1,\ldots,n-1 \nn 
g_i\, g_{i+1}\, g_i &=& g_{i+1}\, g_i\, g_{i+1}  ~~~~~~ \hbox{for} ~~  i = 1,\ldots,n-2\nn 
g_i\, g_j &=& g_j\, g_i\, ~~~~~~~~~~~~~~ \hbox{for} ~~  | i - j | \ge 2 
\eea 

The Hecke algebra has, as a vector space, a basis  $ h (\sigma) $ 
 labelled by the  elements $\sigma$ of $S_n$. This is often called 
 the ``standard basis'' in the literature.
These elements $h(\sigma) $ 
are obtained by expressing the $ \sigma $ as a minimal length word 
in the $s_i$ and then replacing the $s_i$ by $g_i$.  
These Hecke algebras arise as the algebra of operators 
on $ V^{\otimes n } $, the $n$-fold tensor product of the
 fundamental representation of $U(N)$ or $SU(N)$, which commute 
with the action of 
$U_q ( u(N)) $ or $U_q ( su(N) )$, the $q$-deformation of the 
universal enveloping algebra of $u(N)$ or $su(N)$, respectively.
 The action of the 
$q$-deformed enveloping algebras 
  on $ V \otimes V $ is given by the co-product 
$ \Delta $. This
 obeys the following relations with respect to the 
$R$-matrix:
\bea 
\Delta R &=& \Delta^{\prime} R \nn 
(PR) \Delta &=& \Delta (PR) ~.
\eea 
For $h \in U_q $ , if we write $ \Delta ( h ) = h_1 \otimes h_2 $, 
then $ \Delta^{\prime}  ( h ) = h_2 \otimes h_1 $.
 $P$ is the permutation operator. $PR$ is also 
commonly denoted by $ \check R $. For the $R$-matrix 
we will use the conventions of \cite{frt}.  To make that explicit, 
we write $ R_{\sm{FRT}}$. The Hecke algebra is related to the algebra of the 
$\check R_{\sm{FRT}} $ as:
\be\label{grel}  
g = \sqrt {q} \, \check R_{\sm{FRT}} \!\left( {\over}\!\!q_{\sm{FRT}} = \sqrt { q} \right).
\ee
$g_i$ corresponds to $  \check R_{\sm{FRT}} $ acting in the 
tensor product $ V_{i} \otimes V_{i+1} $ and is sometimes called a braid operator. 

Since the centralizer of $U_q$ is the Hecke algebra, we can construct 
the projectors for irreducible representations of $U_q$ in terms 
of words in the $g_i$.  $g_1$
 acts on the product space $V_1\otimes V_2$, therefore there are two
 possible projectors that we can construct \cite{frt}:
\bea\label{N=2proj}
P_{\tableau{2}} &=& { q^{-1} \over q +q^{-1}} ( 1 + q \check R ) = 
 { 1 \over 1 + q }    ( 1 + g ) \nn 
P_{\tableau{1 1}} &=& { q \over q +q^{-1}} ( 1 - q^{-1}  \check R ) = 
 { q \over 1 + q }    ( 1 -  q^{-1} g )~,
\eea 
which project onto the totally symmetric and antisymmetric tensor 
products of the fundamental representation, respectively.
 Using \eq{hecke}, one easily checks that they satisfy
\be\label{proj}
P_R^2=P_R~.
\ee
The symmetric projector is illustrated  in Appendix D 
in terms of properties of  Clebsch-Gordan coefficients 
of $U_q ( su(2) )$.

Projectors are useful to compute characters in a particular
 representation in terms of lower-dimensional representations. 
For example, taking the trace of the above,
\bea\label{simpchar}
\Tr_{\tableau{2}}U&=&{q^{-1}\over q+q^{-1}}
\left((\tr U)^2+q\,\tr\otimes\tr\left(\cR{\over}(U\otimes 1)
(1\otimes U)\right)\right)\nn
\Tr_{\tableau{1 1}}U&=&{q\over q+q^{-1}}\left((\tr U)^2
-q^{-1}\,\tr\otimes\tr\left(\cR{\over}(U\otimes 1)(1\otimes U)\right)\right)~,
\eea
where the traces on the right-hand side are taken in the fundamental
 representation, $\Tr_{\tableau{1}}=\tr_V=\tr$. From now on we will
 indicate such traces by $\tr_n=\tr_{V^{\otimes n}}=
\tr\otimes\ldots\otimes\tr$. The $U$'s in \eq{simpchar}, 
 which are matrix elements 
 of representations of $U_q$ , generate the dual algebra 
to $U_q$ denoted by $ \mbox{Fun}_q(SU(N)) $  or $\mbox{Fun}_q ( U(N))$
( see for example \cite{majid,cosc9807,br} ). 

Using known facts about Hecke algebras and the $q$-deformation
 of the Schur-Weyl duality between $U(N)$ and $S_n$, we will
 now derive the generalization  for arbitrary irreducible 
representations:
\be\label{qprojf}  
P_R = { d_R(q) \over g }\, \sum_{\sigma}q^{ - l( \sigma) }
 \chi_R ( h ( \sigma^{-1}  )  )\, h(\sigma )  ~,
\ee 
where $ l(\sigma )$ is the length of the permutation, i.e
 the number of elements in the minimal presentation of the
 permutation as a product of simple
transpositions. The character is taken in the Hecke algebra $H_n$.
 Without danger of confusion, we will denote $H_n$ and
 $\mbox{Fun}_q (SU(N))$ characters 
with the same symbol.
The characters for low values of $n$ can be read off from the 
tables in \cite{kingwyb,ram}.
$d_R(q)$ is the $q$-deformation of the dimension of a
representation of the symmetric group, and $g$ reduces 
to $n!$ in the classical limit:
\bea\label{dgforms}  
d_R(q) &=&{  \prod_{i\bra j }  ( q^{l_i} - q^{l_j} )\over 
\prod_{i=1}^m (q-1) (q^2 -1) \ldots (q^{l_i} -1 )  }
{  ( q-1)(q^2-1) \ldots ( q^n-1) \over q^ { m(m-1)(m-2)\over 6} }\nn
g &=& {  ( 1 - q ) ( 1-q^2 ) \ldots ( 1 - q^n ) \over ( 1-q )^n }~,
\eea 
where $l_i = \lambda_i + m - i $ and 
$ \lambda_1 \ge  \lambda_2 \ge .. \lambda_m \ge 0 $ are 
the row lengths of the Young diagram, and $m$ is the number of non-zero $\l$'s.

In order to derive \eq{qprojf}, recall the familiar relation
 in the $ q =1 $ case:
\be\label{schurexp}  
\chi_{R} ( U  ) = { 1 \over n ! }  \sum_{\sigma} \chi_{ R } ( \sigma )
\, \tr_n  ( \sigma U ) ~.
\ee 
Here $R$ is both the $U(N)$ reprsentation corresponding to a 
Young diagram and the $S_n$ rep corresponding to the same 
diagram.  The trace on the right-hand side is taken in $ V^{ \otimes n } $, that is
$ U $ acts as $ U \otimes U\otimes ... \otimes U $ and $ \sigma $
 acts by permuting the vectors of the tensor product. 

The above is obtained from the
fact that, if $V$ is the fundamental representation of 
$U(N)$ or the universal enveloping algebra $U (u(N)) $,
 then   $ V^{ \otimes n } $ can be decomposed 
on the terms of the product group $ U(N) \times S_n$ 
as 
\bea\label{tensexp}   
 V^{\otimes n } = \oplus_R   V_R^{U(N)}    \otimes V_R^{S_n}~.
\eea
The sum is over Young diagrams of $S_n$, $V_R^{S_n}  $ is the 
irrep of $S_n$ corresponding to the Young diagram $R$, 
while $  V_R^{U(N)}  $ is the irrep of $U(N)$ corresponding to the 
same Young diagram. Similar relations hold when $U(N)$ is replaced by
 $SU(N)$.   An immediate consequence of the above 
expansion is 
\be\label{multrexp}  
\tr\, ( \sigma U ) = \sum_{ R } \chi_{ R  } ( \sigma )\, \chi_{ R } ( U  ) 
\ee 
Then we can use orthogonality of characters of $S_n$ 
\be 
\sum_{\sigma } \chi_R ( \sigma  ) \,\chi_S ( \sigma^{-1}  )
 = n ! \, \delta_{RS }
\ee 
to obtain (\ref{schurexp}). From (\ref{tensexp}) it also follows that 
\be 
d_R\, \chi_R ( U) = \tr_n (  P_R U ) ~,
\ee 
hence we can read off 
\be\label{classproj}  
P_R = { d_R \over n! } \sum_{ \sigma } \chi_R ( \sigma^{-1} )\, \sigma ~.
\ee

The decomposition analogous to  (\ref{tensexp})
 holds for  $ U_q ( u(N) ) $, when  
$\mathbb{C} S_n$ is replaced by the Hecke algebra  $ H_n ( q ) $ \cite{jimbo}:
\be\label{qtensexp}  
  V^{\otimes n }  = \oplus_{R} V_R^{U_q} \otimes V_R^{H_n} 
\ee 
Here  $V_R^{U_q}$ is the irrep of $U_q ( u(N)) $ corresponding 
to the Young diagram $R$ and $V_R^{H_n}$ is the 
representation of $H_n$ corresponding to the same Young diagram. 
It follows from (\ref{qtensexp}) that
\be\label{qtrschur}  
\tr_{n}\, ( h(\sigma )\, U ) =  \sum_{ R } \chi_R  ( h ( \sigma )    )\,
 \chi_R   ( U ) 
\ee
$U$  lives in the deformed algebra of functions on $U(N)$ denoted 
as $\mbox{Fun}_q ( U(N) )$. This can be defined
 as the dual to $U_q ( U(N) ) $. 
For further discussion on the duality  see  for example  \cite{majid,cosc9807,br}.  
In \eq{qtrschur} $U$ acts as 
\be 
 ( U \otimes 1 \otimes 1 \otimes\cdots ) ( 1 \otimes U \otimes 1\otimes\cdots )
 ( 1 \otimes 1 \otimes U \otimes 1\otimes \cdots ) \cdots 
 ( 1 \otimes 1 \otimes\cdots\otimes 1 \otimes U ) 
\ee  
This product of $n$ $U$'s is dual to the co-product which 
defines the action of $U_q $ on $ V^{\otimes n } $. 

As will be explained in section \ref{sec5} (see also appendix \ref{appD}), quantum traces contain the $u$-element associated to the Hopf algebra 
$ U_q(su(N))$. We get the quantum trace if we take a trace of the action of  $ u\, U $ on the left-hand side of \eq{qtensexp} to get 
\be\label{qqtrschur}  
\tr_n\left(h(\sigma )\, \rho_n ( u \,  U )  \right)
= \sum_{ R } \chi_R  ( h ( \sigma )    )\,
 \chi_R   ( u\, U ) 
\ee 
Here $ \rho_n ( u )  = u^{\otimes n } $ and $ U $ acts as above.    
For the case of diagonal $U$, the formula 
(\ref{qtrschur}) is used in \cite{ram}.

Multiplying the left- and right-hand side of (\ref{qtrschur}) with 
$  \ q^{-l(\sigma)} \chi_S ( h ( \sigma^{-1}  ) ) $, and using the 
 orthogonality relation \cite{gyoja}
for Hecke characters
\be\label{orthogrel}  
\sum_{\sigma } q^{-l(\sigma)} \chi_R (  h(\sigma) )\, \chi_S ( h ( \sigma^{-1}  ) ) 
=  g\,{ d_R(1)  \over d_R(q) }\, \delta_{RS}~,
\ee
we get 
\be 
\sum_{ \sigma } 
q^{-l(\sigma)} \chi_R ( h ( \sigma^{-1}  ) )\, \tr_{n} (   h(\sigma ) U )
= g\, {d_R(1) \over d_R(q) } \,  \chi_R   ( U     )~.
\ee 
This means that the character can be expressed as
\be\label{charexp}  
\chi_R   ( U  )  = {1\over g}
{ d_R(q) \over d_R(1)   } \,
 \sum_{ \sigma } q^{-l(\sigma)}
 \chi_R ( h ( \sigma^{-1}  ) )\, \tr_{n} (   h(\sigma ) U )~.
\ee 
This equation can be interpreted as giving us the 
projection on a fixed Young diagram from the sum in 
(\ref{tensexp}).  Indeed, note that (\ref{qtensexp}) implies,
 by projecting on a fixed Young diagram :  
\be 
d_R(1 )\, \chi_R  ( U  ) = \tr_n   (  P_R   U )  ~.
\ee 
Comparing with (\ref{charexp}) we see that the projector is
\be\label{projform}  
P_{R } = {1\over g}\, d_R(q)\, \sum_{ \sigma } q^{-l(\sigma)}~
 \chi_R ( h ( \sigma^{-1}  ) )\,  h ( \sigma  ) ~,
\ee 
as claimed above. In the appendix we check that it satisfies \eq{proj}.

If we use orthogonality starting from (\ref{qqtrschur}) rather 
than (\ref{qtrschur}), then we get 
\be\label{qcharexp}  
\chi_R^{(q)} ( U )  \equiv  \chi_R   ( u\, U  )  = 
{1\over g}{ d_R(q) \over d_R(1)  } 
 \sum_{ \sigma } q^{-l(\sigma)} \,
 \chi_R ( h ( \sigma^{-1}  ) )\,  \tr_{n} (   h(\sigma )  ( u \, U )  )~.
\ee 
Note that $ u^{\otimes n } $ commutes with 
$h(\sigma )$. We will specialize to $U = 1 $ 
in order to get a new formula for the $q$-dimension in section (\ref{qdimension}).

\subsection{A Hecke formula for the $q$-dimension }\label{qdimension}

Recall that in the case $q=1$ there is a very useful formula 
for the dimension of $SU(N)$ reps which follows from Schur-Weyl
duality \cite{cmri}.
This formula can be obtained by specializing \eq{schurexp} to $U=1$. To that end we need to compute the trace of a permutation acting on $ V^{\otimes n }$. If $ \sigma =1 $, we just 
get $N^n $. If $ \sigma = ( 12) (3) (4) .. (n) $, we get 
$N^{n-1}$. In general we get one factor of $N$ for each cycle in the
permutation. If the permutation has cycles of length $i$ occuring with 
multiplicity $k_i$ the power of $N$ is $N^{ \sum  k_i } $.
In the 2d Yang-Mills literature this is also denoted as $ N^{ K_{\sigma } } $. 
So the useful formula for the dimension in 2d Yang-Mills \cite{gti,gtii}  
is 
\bea 
\mbox{dim}( R ) &=& { 1 \over n ! } \sum_{\sigma } \chi_R ( \sigma )\, N^{ K_{\sigma} } 
 \nn 
%&=&  { 1 \over n ! } \sum_{\sigma } \chi_R ( \sigma ) ~
%N^{ \sum_{i} k_i ( \sigma )   } \nn 
&=& { N^n  \over n ! } \sum_{\sigma  } \chi_R ( \sigma ) \,
N^{ -n + \sum_{i}   k_i ( \sigma )   } ~. \\
&=& { N^n  \over n ! }  \chi_R ( \Omega_n ) 
\eea 
The last line defines the element  $ \Omega_n $. 
It is convenient to write this as a sum over conjugacy classes. 
Let $ T $ be a conjugacy class, which is given by 
specification of the cycle decomposition of the permutations involved. 
We will write $ C_T = \sum_{ \sigma \in T }  \sigma $. 
Note this is a central  element of the group algebra $ \mathbb{C}S_n $, 
i.e.~it commutes with all the elements of $S_n $. So the above can be rewritten as 
\be\label{dimcent} 
\mbox{dim} ( R  ) = { 1 \over n ! } \sum_{ T  } \chi_R ( C_T )\, N^{ \sum_i  k_i ( T ) }~.
\ee    

We can now 
 find the $q$-generalization of this formula by 
setting $U=1$ in  (\ref{qcharexp}), to obtain 
\bea\label{qdimfst} 
\mbox{dim}_q ( R )  =  {1\over g}{ d_R(q) \over d_R(1)} \, \sum_{ \sigma \in S_n  } 
 q^{-l(\sigma) } \,\chi_R \left( h(\sigma^{-1} ) \right) \tr_n ( h(\sigma ) u )~.
\eea 
We can manipulate the above sum, using cyclicity of the trace and the 
Hecke relations, to reduce it to a sum over conjugacy classes $T$ in 
$S_n$, with the only terms appearing inside $ \tr_n $ being 
the $ \tr_{n} (  u \,h ( m_T ) )$.  $m_T $ are permutations 
in the conjugacy class $T$ which have minimal length when expressed 
in terms of generators. They are the minimal words in \cite{kingwyb}. 
 For $n=3$, $ m_T $
 are $ 1, g_1 , g_1g_2 $ for the 3 conjugacy classes. 
We prove in the appendix B (\ref{mintr}) 
\bea\label{truh}
\tr_n \left( u \, h(m_T ) \right)
  = q^{ {N+1\over 2 }\,l(T)}\,  [N]^{ \sum_{i} k_i } 
\eea  
where  the $q$-number is 
\be\label{qnumber}
[N]={ q^{N/2} - q^{-N/2} \over q^{1/2} -  q^{-1/2} }~,
\ee
and $ l(T) $ is the length of the permutation $m_T$.

We will explain below that the Hecke algebra elements  $C_T$ 
appearing as the coefficients of  $ q^{-l(T)} \tr ( u\, h(m_T) ) $  are central. 
Hence the formula for the $q$-dimension becomes 
\bea\label{qdimcent}  
\mbox{dim}_q ( R ) 
%&=&  
%{ 1 \over g }  { d_R(q )  \over d_R(1)  } ~   \sum_{ T } \chi_R 
%  ( C_T ( q) ) [N]^{ \sum_i k_i ( T ) } 
%q^{ ( {N-1 \over 2 }) \sum_i ( i-1) k_i   } \nn
&=&   { 1 \over g } ~  { d_R(q )  \over d_R(1)  } ~\sum_{ T } \chi_R 
  ( C_T ( q) ) ~ [N]^{ \sum_i k_i ( T ) } ~ q^{{N-1 \over 2 }\,l(T)}~.
\eea 
Examples of this formula are described in Appendix B, 
along with  checks against the standard formula  in terms of a product 
of $q$-numbers over the cells of the Young diagram.

We now explain the centrality property of $C_T$.  
Starting from the formula for the projector (\ref{qprojf})
 we can express it in a reduced form using cyclicity and Hecke 
relations, where we 
only have the characters of the minimal words in each conjugacy class: 
\be\label{projmin}  
P_R =  { 1 \over g }\,d_R(q) \sum_{ T } \chi_R ( h ( m_T )  ) \,C_T ~.
\ee 
Here $T$ runs over conjugacy classes, and $m_T $ are 
the minimal words.  For the formulas up to $n=4$, see appendix \ref{appC}.
We can get the projector to the form 
(\ref{projmin}) because, by using 
cyclicity of $ \chi_R $   and the Hecke relations, the Hecke characters can be expressed in terms of these basic 
characters \cite{kingwyb}.  Now for every $R$,  $P_R $
 is   a central element of the Hecke algebra since it is a projector
 for the irreducible representation $R$. 
 There are as many conjugacy classes $T$ as irreducible representations $R$.
 Hence $C_T $ must be central elements. 
When we calculate the $q$-dimension we get  (\ref{qdimfst}).
When we manipulate the expression to express it in terms 
of $ q^{-l(T)}\, \tr_n ( u\, h(m_T)) $, we are using the same Hecke relations 
and cyclicity (of $\tr_n$  this time):
\be 
\mbox{dim}_q (R) =   { 1 \over g }\,{d_R(q)\over d_R(1)} \,\sum_{ T }q^{-l(T) }\, \chi_R ( C_T    ) \,
\tr_n  ( h ( m_T )\,  u )~.  
\ee 
 This immediately leads to (\ref{qdimcent}). Incidentally,
 \eq{projmin} seems to give a relatively efficient way of caculating 
the central  class elements compared to the ones
 we are aware of in the mathematical literature.
Some  interesting papers with explicit formulae for 
 Hecke central elements, which we found useful,
 are \cite{francis99,francisjones}.

\subsection{Hecke  $q$-generalization of sums 
over symmetric groups of 2d Yang Mills } 

The string theory interpretation of 2d Yang Mills at $q=1$
is centred on formulae derived from Schur-Weyl duality. 
The character relations following from Schur-Weyl 
give rise to a formula for dimensions of $SU(N)$ reps 
in terms of $S_n$ reps. Then some group theory manipulations lead 
to an expression of the chiral partition function in terms 
of delta functions over the symmetric group. 

The delta function is defined over the symmetric group 
or, more generally, over the group algebra of the symmetric group:
\bea 
\delta ( \sigma ) &=& 1  ~~~ \hbox{if} ~~ \sigma = 1  \nn 
\delta ( \sigma )  &=& 0 ~~~ \hbox{otherwise}~. 
\eea 
A useful property of this delta function is that it can be expressed 
in terms of characters,
\be\label{delchar}  
n!\,\delta ( \sigma) = \sum_{ R } d_R\, \chi_R (\sigma)~. 
\ee 

The expressions arising in the 2d Yang-Mills string  take the form 
\be 
\delta  \left(\sigma_1 \sigma_2 \cdots \sigma_k\right)~,
\ee
and the weights depend on the genus $G$ and on  $N$ 
in precisely such  way that the chiral partition 
function can be expressed in terms of a sum of Euler 
characters of moduli spaces of holomorphic maps 
(see section 7 of \cite{cmri}).  

Now we will  describe  a $q$-generalization of this story, 
where the Hecke algebra will replace the group
algebra of the symmetric group. A  $q$-analog of the 
delta function on the symmetric group is known in the theory 
of Hecke algebras \cite{gyoja}. 
%See beginning of section 3 there where it is denoted as 
%$ \tr ( h(w) ) $. 
It is defined as:
\bea 
\delta (h( \sigma )) &=& 1  ~~~ \hbox{if} ~~ \sigma = 1  \nn 
\delta (h( \sigma ))  &=& 0 ~~~ \hbox{otherwise}~.
\eea 
Our $ \delta  ( h( \sigma )) $ is $ { 1\over g }\, \tr ( h(\sigma) ) $ 
in the notation of  \cite{gyoja} for the canonical trace function $ \tr ( h(\sigma) ) $. 
This $q$-deformed delta function  reduces exactly to the delta function 
on the symmetric group defined above when $q \rightarrow 1 $. 
%So we will write it as $ \delta $ and our goal will be to 
%write the q2D-Yang Mills partition function in terms of things like 
%\be
%\delta (h(\sigma_1 )   h ( \sigma_2 )\cdots) 
%\ee 
It can be expressed as
\be\label{qdelchar} 
g\, \delta (  h( \sigma ) ) = \sum_{ R }  d_R(q)\,\chi_R  ( h(\sigma) ) ~,
\ee 
where $R$ runs over partitions of $n$ or Young diagrams with $n$ boxes, 
and $g$ is given in (\ref{dgforms}).

An important fact we will use in what follows is that
 for $C$ a central element of the Hecke algebra, and for arbitrary $\s\in S_n$,
\be \label{characterfusion}
\chi_R (C)\, \chi_R \left(h(\s)\right) = d_R(1)\,  \chi_R \left(C\,h(\s)\right) ~,
\ee 
which follows simply from Schur's lemma applied to the Hecke algebra.

We now have all the elements we need in order to rewrite the quantum dimensions in terms of central elements of the Hecke algebra. Using (\ref{qdimcent}), we can write
\be\label{qdim1}
\mbox{dim}_q ( R ) = { [N]^{ n } \over g } { d_R(q) \over d_R(1) }\,
   \chi_R (  \Omega_n )~.
\ee
In the quantum case the $\Omega$'s are expressed as
\bea
\Omega_n   &=& \sum_{T} [N]^{K_T - n } \,q^ { {  N-1\over 2 }\,l(T)}\, C_T  \nn 
% & =& 1 + { 1 \over [N] } q^ { {  N-1\over 2 }  } \chi_R ( C_{(2)} ) 
%        + { 1  \over [N]^2 } q^{ 2 {  N-1\over 2 }  } \chi_R ( C_{(3)} ) 
%        +  { 1  \over [N]^2 } q^ { 2 {  N-1\over 2 }  }\chi_R ( C_{(2,2)} )\nn 
& =& 1 + \sum_{T}{}'~
%^{~~~~~\prime} ~~
  [N]^{K_T - n } \,q^ { {  N-1\over 2 }\, l(T) }\,
    C_T  \nn 
&  \equiv&  1 + \Omega'_n  ~,
\eea 
where the unprimed sum runs over the central elements of $H_n$. 
%The characters $\chi_R $ are characters of the Hecke representation $R$. 
The restricted sum (denoted by the prime) runs over all central elements
associated with conjugacy classes of $S_n$ 
which are not the identity. The last line is a definition 
of  $\Omega^{\prime}_n $. 
Making repeated use of \eq{characterfusion}, we find that for a central element we have:
\be 
\left(  { \chi_R (C )  \over d_R(1 )  } \right)^m =  { \chi_R (C^m  )  \over d_R(1 )  }~.
\ee 
It now follows from \eq{qdim1} that
\bea\label{qdimm}
  ( \mbox{dim}_q  ( R ) )^m \!\!\!\! &=& \nn 
%& =&  \left( { [N] d (R,q )   \over g } \right)^m   
%\left(  { \chi_R ( \Omega ) \over d (R,1 )   }  \right)^m \nn 
\!\!\!\!& =& \!\!\!\!  \left( { [N]^n\, d_R(q)  \over g } \right)^m  
  { \chi_R ( \Omega^m ) \over d_R(1 )  }  \\
%& =&  \left( { [N]^n d_R(q) \over g }  \right)^m 
% \sum_{\ell=0}^{\infty } d ( m , \ell ) 
% ( \Omega^{\prime} )^\ell  \nn 
\!\!\!\!  &=&\!\!\!\! \left( { [N]^n\, d_R(q)  \over g }  \right)^m    \sum_{\ell=0}^{\infty } {d ( m , \ell ) \over d_R(1)}
\chi_R \left(\prod_{i=1}^\ell \sum_{T_i}{}' C_{T_i}\right)[N]^{\sum_iK_{T_i}-n} q^ { {  N-1\over 2 }\sum_{i} l ( T_i )   }\nonumber
%&\times&   \sum_{T_1}{}' \sum_{T_2}{}' \cdots \sum_{T_L }{}'\,
% C_{T_1} C_{T_2} \cdots C_{T_L} 
% [N]^{  \sum_i K_{T_i}  - n } \,q^ { {  N-1\over 2 }\, ( \sum_{i} l ( T_i )  ) }~.
\eea 
where $ d ( m , \ell ) = {  \Gamma ( m +1 ) \over \Gamma ( \ell+1 ) \Gamma ( m-\ell+1 ) } $, and we wrote out  the definition of $\O'$.

Let us develop the $q$-deformed chiral Gross-Taylor expansion
\bea\label{qcgt1}  
Z &=& \sum_{ n = 0 }^{\infty } \sum_{ R \in Y_n } 
       \left( \mbox{dim}_q  ( R ) \right)^{ 2 - 2G }   \nn 
& = &\sum_{ n = 0 }^{\infty } \sum_{ R \in Y_n } 
    [N]^{(2-2G)n}   \left({  d_R(q) \over g }\right)^{2-2G} \, 
%\left( { g^2 \over d_R(q )} \right)^{G} 
    { 1\over d_R(1) }\,\chi_R \left( \Omega^{ 2 - 2G } \right)  ~.
\eea 
 Now we can show  (see appendix A) that
\bea\label{qdefpirel}  
\left( { g \over d_R(q ) } \right)^2 =\sum_{s,t\in S_n}  
 q^{ - l(s) -l(t) }\,{ 1 \over d_R(1)  }\, \chi_R   \left( h(s) h(t) h(s^{-1})h( t^{-1} )   \right)~.
\eea 
We also show in the appendix that the element
\be\label{cqdefpirel}
\sum_{s,t\in S_n}  q^{ - l(s) -l(t) } \,  h(s) h(t) h(s^{-1} ) h( t^{-1} ) 
\ee 
is central in $H_n$.  Hence we have 
\bea\label{expnfac} 
\left( { g \over d_R(q) } \right)^{2G}=\sum_{s_1, t_1  \cdots s_G,t_G }   q^{ -\sum_i  ( l(s_i ) + l(t_i ) )}\,
{1 \over d_R(1) }\,\chi_R  \left(  \prod_{i=1}^G h(s_i) h(t_i) h(s_i^{-1} ) h( t_i^{-1} ) \right)~.
\eea 
Now we employ this equation in  (\ref{qcgt1}) to get 
\bea\label{qcgt2}  
Z &=& \sum_{ n = 0 }^{\infty } \sum_{ R \in Y_n }  \sum_{s_it_i} 
   q^{ -\sum_i  ( l(s_i ) + l(t_i ) ) }\,
    [N]^{(2-2G)n}   \left({  d_R(q) \over g\,d_R(1) }\right)^2\! \times \nn
&\times& \, \chi_R  \left(\prod_{i=1}^{G}h(s_i) h(t_i) h(s_i^{-1} ) h( t_i^{-1} )\right)
    \chi_R ( \Omega^{ 2 - 2G } )  \nn 
& =&  \sum_{ n = 0 }^{\infty } \sum_{ R \in Y_n } \sum_{s_it_i}\,  
   [N]^{(2-2G)n}  \left({  d_R(q) \over g }\right)^2 {q^{ -\sum_i  ( l(s_i ) + l(t_i ) ) }\over d_R(1)}\, \times \nn
&\times& \, \chi_R \!\left(\Omega^{ 2 - 2G } \prod_{i=1}^{G}   h(s_i) h(t_i) h(s_i^{-1} ) h( t_i^{-1} ) \right)
\eea 
where we sum over $S_n$ permutations $s_1,t_1,\ldots,s_G,t_G$.
At this point the manipulations performed in the classical case
 do not generalize straightforwardly to the quantum case because of the
different powers of $d_R(q)$ and $d_R(1)$.
We need to introduce an  element $D$ of the Hecke algebra with the 
property 
 \be \label{Delement}
\chi_R ( D ) = d_R(q )~. 
\ee 
The existence of this element is proven in appendix \ref{centralelements}, where an explicit expression is given for it in terms of an
infinite sum. Let us find it explicitly
for low values of $n$. For $n=2, 3 $, we can solve the above equation explicitly. We find for $n=2$ 
\bea 
D &=& {  1 + q^2  \over 1+q   } +  {  1-q \over  1+q  }\, g_1    \nm
\eea 
and for $n=3 $ 
\bea
D &=&   { 1 + q^2+2q^3+q^4+q^6 \over (1+q)(1+q+q^2)}\, +  
{ (1-q)( 2+2q+ q^2 + 2q^3+2q^4) \over ( 1+q ) ( 1+q +q^2)  }\, g_1  \nn 
 &&+ {  ( 1+q) (1-q)^2 \over 1+q+q^2 }   \,  g_1 g_2~.
\eea 
%A general description of $D$ is given as an infinite sum 
%in Appendix A. 
We note that $D\rightarrow1$ in the classical limit.
Using the form of $D$ in the appendix we can write $ { d_R(q)^2 \over d_R(1 ) }  = 
d_R(q)\,  { \chi_R ( D ) \over d_R(1) } $, which allows us to rewrite 
(\ref{qcgt2}) 
\bea\label{qcgt3}  
Z   &=&  \sum_{ n = 0 }^{\infty }
 \sum_{s_it_i }  
   {  1 \over g }\,[N]^{(2-2G)n}\,   \delta \left(D \,  \Omega^{ 2 - 2G } \prod_{i=1}^{G} q^{-l(s_i) -l(t_i) } 
   h(s_i) h(t_i) h(s_i^{-1} ) h( t_i^{-1} ) \right).
\eea 
In the last step we used (\ref{qdelchar}).  
This is the q-analog of the Gross-Taylor  expansion.
We can expand the $\O$-factors as follows
\bea\label{qcgt4}
Z&=& \sum_{ n = 0 }^{\infty }  \sum_{s_it_i}q^{ -\sum_i  ( l(s_i ) + l(t_i ) ) } \sum_{\ell=0}^{\infty }  
        \sum_{ T_1\ldots T_\ell }\!\!\!{ \over }^{\prime}\,
 {1\over g}\, [ N ]^{(2-2G)n+ \sum_i   ( K (T_i) - n )  }\, q^{ {N-1 \over 2} \sum_{ i=1}^\ell l(T_i )} 
     \nn 
&\times&   d\left(2-2G , \ell \right)  \,
 \delta \left( D \, C_{T_1} \ldots C_{T_\ell }   \prod_{i=1}^{G}   h(s_i) h(t_i) h(s_i^{-1} ) h( t_i^{-1} ) \right) .
\eea  
As explained  in \cite{cmri}, the factor of $  d ( 2-2G , \ell ) $ 
is the Euler character of the configuration space 
of $\ell $ points on $ \Sigma_ G $, denoted as 
  $ \chi  ( \Sigma_{G , \ell}  )$. Hence we can write 
\bea 
Z  &= &
\sum_{ n = 0 }^{\infty } \sum_{s_it_i}\,  \,  q^{ -\sum_i  ( l(s_i ) + l(t_i ) ) }
\sum_{\ell=0}^{\infty}  
\sum_{ T_1\ldots T_\ell }\!\!\!{ \over }^{\prime} \,{1\over g}\,[ N ]^{(2-2G)n+ \sum_{j=1}^{\ell}   ( K (T_j) - n )  }\,q^{ {N-1\over 2}\, \sum_{ j=1}^\ell l(T_j)}\nn 
&\times& \chi  ( \Sigma_{G ,\ell}  )\, ~~  \d\left(  D \, C_{T_1} \ldots C_{T_\ell }   \prod_{i=1}^{G} q^{ - ( l(s_i ) + l(t_i ) ) }  h(s_i) h(t_i) h(s_i^{-1} ) h( t_i^{-1} )\right). 
\eea

\section{Manifolds with boundary } 

We now describe  the chiral large $[N]$ expansion of  $q$-deformed
 2d Yang-Mills theory on manifolds with boundary,  in terms of Hecke algebras. 
 We  recall the classical case first. For a Riemann surface of 
genus $G$ with $B$ boundaries and boundary holonomies $U_1,\ldots, U_B $ in $SU(N)$, the parition function is 
\be
 Z_{\sm{YM}}( G , B ; U_1,\ldots, U_B ) 
= \sum_{R }  ( \mbox{dim}\, R )^{2-2G - B } \chi_R ( U_1)\, \chi_R ( U_2) \ldots \chi_R ( U_B )~.  
\ee 
It is useful in that case to multiply by  
 $ ( { 1 \over n! }  )^B\, \tr_n  ( T_1\,  U_1^{\dagger}  )\, \tr_n ( T_2\, U_2^{\dagger}  ) \ldots
 \tr ( T_B\,  U_B^{\dagger}   ) $
and integrate over the holonomies, where $T_1,\ldots,T_n $ are sums of 
permutations in fixed   
conjugacy classes in $S_n$. Then the chiral Gross-Taylor expansion becomes 
\be 
Z_{\sm{YM}}  ( G , B ; T_1,\ldots, T_B  ) 
= \sum_{ s_i,t_i } { 1 \over n! }\,
 N^{ n ( 2 - 2 G - B ) } \delta \left(  T_1 \ldots T_B\, 
\Omega_n^{2-2G -B }  \prod_{i=1}^{G}  s_i t_i s_i^{-1} t_i^{-1}  \right)~.
\ee 
This is basically a Fourier transformation, and the  derivation is explained in \cite{ramwilson}.

For $q$-deformed 2d Yang-Mills, the holonomies along the boundaries are specified by the quantum characters \cite{br,boulatov} of
$U_q ( SU(N)) $:
\be
 Z_{\sm{$q$YM}}( G , B ; U_1,\ldots, U_B ) 
= \sum_{R }  ( \mbox{dim}_q\, R )^{2-2G - B } \chi_R ( U_1)\, \chi_R ( U_2) \ldots \chi_R ( U_B )~.
\ee 
Now we can insert 
 $ ( { 1 \over g}   )^B\,  \tr_n  ( C_{T_1}\,  U_1^{\dagger}  )\, \tr_n ( C_{T_2 }\,  U_2^{\dagger}  )\ldots
 \tr ( C_{ T_B }\,   U_B^{\dagger}   ) $. In this case, $ C_{T_1}, \ldots,  C_{T_B} $ are central elements
in $H_n(q) $ which approach the class sums $T_1, T_2,\ldots, T_B$ 
in the limit $ q \rightarrow 1$. They have appeared in the formulae
 for the $q$-dimension earlier. 
 We use the expansion 
\be 
\tr ( C_{T}\, U^{\dagger} ) = \sum_{ S }  \chi_S ( C_T )\, \chi_S ( U^{\dagger} )~, 
\ee 
 where  $ \chi_S ( C_T ) $ is the Hecke algebra character in the representation 
$S$. 
Then we integrate the quantum group elements $ U_1,\ldots, U_B $, and use 
the orthogonality \cite{br,boulatov}
\be \label{glue}
\int\dd U\, \chi_R ( U )\, \chi_S ( U^{\dagger} ) 
= \delta_{ RS } ~.
\ee 

The result is 
\bea\label{bdyres}  
&Z_{\sm{$q$YM}}&\!\!\!\!\!( G , B ; C_{ T_1 },\ldots, C_{ T_B }   ) 
= \sum_{ R \in Y_n }  \left( \mbox{dim}_q\,  R\right)^{ 2 - 2G - B } \, \prod_{j=1}^{B}
 \left( { \chi_R ( C_{T_j} ) \over g  }\right)   \nn
& =&\!\!\!\!\!\!\!\! \sum_ { R }\,[N]^{  (2-2G-B)n } 
\left( { d_R(q) \over g }\,   {\chi_R ( \Omega ) \over
 d_R(1 )  } \right)^{ 2-2G - B } \prod_{j=1}^{B} \left(\chi_R ( C_{T_j} )\over g\right) \\
%&=& [N]^{ n (2-2G-B) }
%\sum_ { R } ({ d_R(q)\over g })^2   ( { g^2 \over d_R(q)^2})^{G } 
% { \chi_R  \over d ( R,1) } \bigl ( \Omega^{2-2G-B } \bigr )
% { d( R,q)}^{-B }   \prod_{j=1}^{B} \chi_R ( C_{T_j} )   \nn
%&=&[N]^{ n (2-2G-B) } \sum_R  ( { d ( R,q) \over g } )^2  \sum_{s,t}  
%(~  { \chi_R \over d(R,1 )}  ( q^{-l_s} h(s) h(t) h(s^{-1} )
% h(t^{-1} ) )~ )^{G }
% \nn 
%&&  \qquad \qquad  \qquad \qquad 
%{ \chi_R \over d_R(1) }  ( \Omega^{2-2G -B } ) 
% ( { d_R(1) \over d_R(q) })^{B }  { \chi_R \over d_R(1) }
% ( C_{T_1} .. C_{T_B} )  \nn
%&=& [N]^{ n (2-2G-B) } \sum_R   ( { d_R(q) \over g })^2
%  ( { \chi_R \over d_R(1) } ( { E\over g } ))^B  \sum_{s_i,t_i } \nn 
%&& \qquad \qquad 
% { \chi_R \over d_R(1) } (  \Omega^{2-2G - B } \prod_{i=1}^{G}
% q^{-l(s_i) - l(t_i) } h(s_i)h(t_i) h(s_i^{-1})h( t_i^{-1} ) 
% \prod_{j=1}^{B } C_{T_j} ) \nn 
%& =&  [N]^{ n (2-2G-B) } \sum_R   ( { d_R(q) \over g })^2 \sum_{s_i,t_i }  \nn 
%&& \qquad 
%   { \chi_R \over d_R(1) } ( ~  ( { E\over g } )^B 
%\Omega^{2-2G - B } \prod_{i=1}^{G}
% q^{-l(s_i) - l(t_i) } h(s_i)h(t_i) h(s_i^{-1})h( t_i^{-1} ) 
% \prod_{j=1}^{B } C_{T_j} ~ )  \nn
%& =& [N]^{ n (2-2G-B) } 
% \sum_R  ( { d_R(q)\over g^2 }) { \chi_R } ( gE^{-1}  ) \sum_{s_i,t_i }  \nn 
%&& \qquad 
%       { \chi_R \over d_R(1) } (~  ( { E\over g } )^B 
%\Omega^{2-2G - B } \prod_{i=1}^{G}
% q^{-l(s_i) - l(t_i) } h(s_i)h(t_i) h(s_i^{-1})h( t_i^{-1} ) 
% \prod_{j=1}^{B } C_{T_j} ~ )  \nn 
& =&\!\!\!\!\!\!\!\!   {1\over g}\,[N]^{ (2-2G-B)n }  \sum_{s_it_i } \,
 \delta\left(  \left({ E\over g }\right)^{B-1}  \Omega^{2-2G - B }
\prod_{i=1}^{G} q^{-l(s_i) - l(t_i) }\, h(s_i)h(t_i) h(s_i^{-1})h( t_i^{-1} ) 
 \prod_{j=1}^{B } C_{T_j} \right). \nonumber
\eea     
In the second line we used \eq{qdimm},
and in the last line we employed  \eq{expnfac}. 
The element $E$ is defined in (\ref{defE}). 
As in manipulations of the partition 
function we repeatedly used 
 \eq{characterfusion} to  combine products of characters.  
 Finally to obtain the delta function from the Hecke characters, we used 
 \eq{qdelchar}.

In the $q=1$ limit (\ref{bdyres})  reduces to  a
 delta function over the group algebra of $S_n$,  
counts maps with specified conjugacy classes of 
 permutations at the boundaries. 
There is now some deformation of this geometry, involving 
central elements of the Hecke algebra $H_n(q)$ associated 
 with the boundaries. It is very intersting that for $B=1$
 we do not have the ${ E \over g }  $ factors.
 Recall also that $  { E\over g }  =1$ in the $q=1$ limit.

In the $q$-deformed theory there is a notion of a delta-function over the
quantum group -valued holonomies \cite{br}.
It is the partition function on the disk, therefore the case $G=0$, 
$B=1$ of the above. We compute directly:
\be
\d(U,1)=\sum_{n;\s\in S_n}\,{1\over g}\,[N]^n\,q^{-l(\s)}
\d( D  \O\, h(\s^{-1}  ))\,\tr_n(h ( \s ) \,u\,U)=\sum_{n;\s\in S_n}\,
{1\over g}\,[N]^n\,Q^\s\,\tr_n(h ( \s ) \,u\,U)~,
\ee
where we defined $D \O=\sum_\s  Q^{\s } h(\s)$.
 Using \eq{glue}, we can integrate this expression against 
any test function to obtain
a form that depends purely on the Hecke algebra. In particular, 
the above gives another expression for the quantum dimensions. Thus,
 in the $q$-deformed theory the partition function on a 
disk of zero area continues to be associated to
a flat connection, in the quantum group sense \cite{br}.

\section{Chiral large $N$ expansion for Wilson loops } 

After having computed the partition function on closed Riemann surfaces and Riemann surfaces with boundaries, we should now
discuss the chiral expansion of Wilson loops. For simplicity, we will consider non-intersecting Wilson loops in this section.
The basic object we need to take into account are the $SU(N)$ tensor multiplicity coefficients \cite{br,boulatov}. Indeed, consider 
a surface of genus $G=G_1+G_2$ with a Wilson loop in representation $S$, where $G_1$ and $G_2$ are the genera of the inner and outer 
faces of the Wilson loop. The expectation value of this Wilson loop is
\be
W_S(G)=\sum_{R_1R_2}\int\dd U\,\left(\mbox{dim}_q\,R_1\right)^{1-2G_1}\left(\mbox{dim}_q\,R_2\right)^{1-2G_2}\,\chi_{R_1}(U)\,\chi_{S}(U)\,
\chi_{R_2}(U^\dagger)~,
\ee
where $R_1$ and $R_2$ are the representations of the inner and outer faces, respectively. Since we are discussing the case of $q$ non-root of unity, the result of the above quantum integral are the usual $SU(N)$ tensor
multiplicity coefficients (Littlewood-Richardson coefficients). Thus we are set to compute
\be\label{wilsonloop}
W_S(G)=\sum_{R_1R_2}(\mbox{dim}_qR_1)^{1-2G_1}(\mbox{dim}_qR_2)^{1-2G_2}\,N^{R_2}_{R_1S}~.
\ee

Our next task is to look for an expression for the Littlewood-Richardson coefficients that we can interpret as a deformation of the 
Riemann surface. Thus, we want to write them as delta functions on the Hecke algebra. We start from the definition:
%Consider the simple case of a cylinder, with boundary conditions 
%restricting the boundary holonomies to take values $U_1,U_2$
% at the two boundaries, and we insert a Wilson loop 
\bea 
N_{R_1 R_2}^{R_3} = \int \dd U\, \chi_{R_1} ( U )\, \chi_{R_2} ( U )\,  \chi_{R_3} ( U^{\dagger} ) 
\eea 
and observe that the above is a trace of the following operator
acting in $ R_1 \otimes R_2 $:
\bea 
  \int \dd U\, \chi_{R_3} ( U^{\dagger} ) \, \rho_{R_1 \otimes R_2 } ( U ) ~.
\eea 
Now $R_1$ can be realized in $ V^{\otimes n_1} $ with multiplicity 
$d_{R_1}(1 )$ when we project on the given Young diagram, and likewise for 
$R_2$. It is also useful to note that the above operator is 
proportional to a projector for the representation $R_3$
\bea 
  \int \dd U\,  \chi_{R_3} ( U^{\dagger} )\,  \rho_{R_1 \otimes R_2 } ( U )  
= { 1 \over \mbox{dim}_q\, R_3  }\,  \rho_{R_1 \otimes R_2 } ( P_{R_3} )~.
\eea 
Using the expression for the projectors  for $R_1$ and $R_2$ 
in terms of the Hecke algebra, we obtain 
\bea
N_{R_1 R_2}^{R_3} &=& {1\over g_1g_2g_3}\, {d_{R_1}(q)\over d_{R_1}(1)}\,{d_{R_2}(q)\over d_{R_2}(1)}\, \sum_{ \sigma_1\sigma_2 } 
  q^{-l(\s_1)-l(\s_2) }  \, \chi_{R_1} ( h(\s_1^{-1} ) ) \, 
 \chi_{R_2} ( h(\s_2^{-1} ) ) \nn
&\times& { 1 \over \mbox{dim}_q R_3 }\, \tr_{V^{\otimes n_1} \otimes V^{\otimes n_2}} \left( ( h(\s_1)\cdot h(\s_2) ){\over} P_{R_3} \right).
\eea
Here and in what follows we take $\s_i\in S_{n_i}$ for $i=1,2,3$.
Writing out the projector \eq{projform}, we get
\bea
N_{R_1R_2}^{R_3}&=& {1\over g_1g_2g_3}\, {d_{R_1}(q)\over d_{R_1}(1)}\,{d_{R_2}(q)\over d_{R_2}(1)}\, \sum_{\s_1\s_2\s_3} 
q^{-l(\s_1)-l(\s_2)-l(\s_3) }\,\chi_{R_1} ( h(\s_1^{-1} ) )\,
\chi_{R_2} ( h(\s_2^{-1} ) )\,
\chi_{R_3} ( h(\sigma_3^{-1}  ) ) \nn
&\times&{ d_{R_3}(q) \over \mbox{dim}_q R_3 }\, \tr_{V^{\otimes n_1} \otimes V^{\otimes n_2} }\left( ( h(\s_1)\cdot h(\s_2)){\over}h(\s_3)\right),
\eea
and expanding the trace in a basis of Young tableaux with
 $n_1+n_2$ boxes, we get
\bea\label{fus} 
N_{R_1R_2}^{R_3}&=& {1\over g_1g_2g_3}\, {d_{R_1}(q)\over d_{R_1}(1)}\,
{d_{R_2}(q)\over d_{R_2}(1)}\, \sum_{ \sigma_1 \s_2\s_3} 
q^{-l(\sigma_1) - l(\sigma_2) -l(\sigma_3 ) } \,
 \chi_{R_1} ( h(\s_1^{-1} ) )\, \chi_{R_2} ( h(\s_2^{-1} ) )\,  \chi_{R_3}
 ( h(\sigma_3^{-1} ) )\nn               
&\times& {d_{R_3}(q) \over \mbox{dim}_q R_3 }\, \sum_{S\in Y_{n_1+n_2} } 
\chi_S \left(( h(\s_1)\cdot h(\s_2)){\over}h(\s_3)\right)\mbox{dim}_q\,  S~.
\eea 
If we now use the projector property
\be 
\chi_S ( P_{R_3} h(\sigma)  ) = \delta_{R_3 S } \,
{ \chi_{R_3} (  h(\sigma) ) }
\ee
and the explicit form of the projector in (\ref{qprojf}) then we have   
the  useful  orthogonality relation 
\be 
\sum_{\sigma_3 } q^{-l(\sigma_3)}\,   {d_{R_3}(q) \over g_3}\,
\chi_{R_3} ( h(\sigma_3^{-1} ) ) \,
\chi_S \left({ \over } h(\sigma_3)  ( h(\sigma_1)\cdot h(\sigma_2) )\right)  
=  { \chi_{R_3}  ( h(\sigma_1)\cdot h(\sigma_2) )  } \,
\delta_{R_3 S }
\ee 
This can be used 
to simplify the expression (\ref{fus}) further to:
\bea \label{tensorprod}
N_{R_1, R_2}^{R_3} = {1\over g_1g_2}\, {d_{R_1}(q)\over d_{R_1}(1)}{d_{R_2}(q)\over d_{R_2}(1)}\,
\sum_{\sigma_1\sigma_2} q^{-l(\sigma_1) - l(\sigma_2) }\, 
\chi_{R_1} ( h(\sigma_1^{-1} ) ) \,\chi_{R_2} ( h(\sigma_2^{-1} ) )\,
 \chi_{R_3} ( h(\sigma_1)\cdot h(\sigma_2)) \nn 
\eea 
This formula is reminiscent of the 
Verlinde formula for the fusion coefficients of orbifold 
conformal field theories \cite{dvvv}, or alternatively of
 Chern-Simons theory with
finite groups \cite{dw,fq}. It would be interesting to understand the connection.

If we go from the character basis to the basis in terms of 
central elements of the Hecke algebra, and using the above,  we get 
\bea 
&& {1\over g_1g_2g_3}\sum_{R_1R_2 R_3} N_{R_1R_2}^{R_3}\, 
\chi_{R_1} \!(C_1 )\, \chi_{R_2} ( C_2 ) \,\chi_{R_3} ( C_3 ) \nn 
&&=  {1\over g_1g_2}\,\sum_{\sigma_1 \sigma_2 }\,  \delta ( h(\s_1^{-1} ) C_1 )\,
\delta ( h(\s_2^{-1} ) C_2 )\,q^{-l(\sigma_1) - l(\sigma_2) }  \,
\sum_{R_3} { 1 \over g_3 }\, d_{R_3}(1) \,
 \chi_{R_3} ( h(\sigma_1)\cdot h(\sigma_2)  C_3 )    \nn 
&& =   {1\over g_1g_2}\,\sum_{\sigma_1 \sigma_2 }\,  \delta ( h(\s_1^{-1} ) C_1 )\,
\delta ( h(\s_2^{-1} ) C_2 )\,q^{-l(\sigma_1) - l(\sigma_2) }  \,
  { 1 \over g_3 }\, \delta \left( E\, C_3 {\over} ( h(\sigma_1)\cdot h(\sigma_2)) \right) \nn 
&& =  { 1 \over g_1 g_2 g_3 }
 \, \sum_{\sigma_1 } \sum_{\sigma_2}
      C_1^{\sigma_1} \, C_2^{\sigma_2}\,
  \delta \left({\over}  E \, C_3 \,    ( h  ( \sigma_1  ) \cdot h  (
 \sigma_2 )) \,  \right).
\eea
$E$ is the element defined in (\ref{defE}) of Appendix A. 
We have denoted by $C^{\sigma } $ the coefficients 
which appear in expansion of the central element $C$, 
\be 
 C = \sum_{\sigma }  C^{ \,\sigma } \, h(\sigma   )  ~,
\ee 
and we used the following property of the trace \cite{gyoja}:
\bea
\d(h(\s)h(\s'))&=&q^{l(\s)}~~~~\mbox{if}~~~\s\s'=1\nn
\d(h(\s)h(\s'))&=&0~~~~~~~~\mbox{otherwise}.
\eea

Consider now the computation of a simple Wilson loop, in the 
representation $S$,   separating  
a region with $G_1$ handles from another region with 
$G_2$ handles. 
\bea 
 W_S  &=&
\sum_{n_1, n_2 } 
 \sum_{R_1R_2} \,  ( \mbox{dim}_q\, R_1 )^{1-2G_1}  ( \mbox{dim}_q\, R_2 )^{1-2G_2}
 N_{R_1 S}^{R_2}  \nn  
& =& \sum_{R_1}\, [N]^{n_1(1-2G_1)}  
\left( { d_{R_1}(q)\over g_1 d_{R_1}(1)  }\right)^{1-2G_1} 
(\chi_{R_1} ( \Omega ))^{1-2G_1}    \nn 
&\times&  \sum_{R_2}\, [N]^{n_2(1-2G_2)}  
 \left( { d_{R_2}(q)\over g_2 d_{R_2}(1)  }\right)^{1-2G_2} 
  (\chi_{R_2} ( \Omega ))^{1-2G_2}    ~~ N_{R_1 S }^{R_2}  ~~ 
\eea 
We now use (\ref{tensorprod}) with the fusion coefficient,  
multiply  by the character of some central element $C$  
in $H_{n_S}(q)$ and sum over $S$ 
\bea 
&& W ( C , G_1, G_2  ) = \sum_S { \chi_{S} ( C ) \over g_S }\, W_S  
\eea  
Collecting all $S$ dependences we have 
\be
  \sum_S \,{1\over g_S}\,{d_S(q) \over d_S(1)  }\, \chi_{S} ( C )\,  \chi_{S} ( h ( \sigma_2^{-1}  ) )   =
 \delta ( C\, h(\sigma_2^{-1}  ) ) 
\ee
Hence we obtain 
\bea 
 W(C; G_1 , G_2 ) &=& \sum_{ n_1, n_2  } \,
  { 1 \over {g_1 g_2 }} \, \delta_{n_1 + n_S , n_2 } \,
   [N]^{n_1 ( 1-2G_1) +n_2 ( 1-2G_2 )} \sum_{\sigma_1 \sigma_2} q^{-l(\sigma_1) - l(\sigma_2 )} \,\times\\ 
&\times&  \delta ( C\, h(\sigma_2^{-1} ) )\,
\delta \left( D\,  \Pi_1^{G_1} \, \Omega^{1-2G_1}\,h ( \sigma_1^{-1}  ) \right)
\delta \left(  \Pi_1^{G_2} \Omega^{1-2G_2}   ( h ( \sigma_1) \cdot h ( \sigma_2 ) ) \right)\nonumber
\eea  
The factors of $[N] $ are as above. We have defined 
\bea 
&& \Pi_1^{G_1} = \sum_{s_1,t_1 .. s_{G_1},t_{G_1}  } 
q^{- \sum_i l(s_i) -l(t_i) } \prod_{i=1}^{G_1} h(s_i) h(t_i) h(s_i^{-1} ) h(t_i^{-1})  \nn
&&  \Pi_1^{G_2} =   \sum_{s_1,t_1 .. s_{G_2},t_{G_2}  } 
q^{- \sum_i l(s_i) -l(t_i) }  \prod_{i=1}^{G_2}
   h(s_i) h(t_i) h(s_i^{-1} ) h(t_i^{-1}) 
\eea 
Expanding
\bea
C&=&\sum_\s C^{\,\s }\,h(\s  )\nn
P &=&D\,\O^{1-2G_1}   \Pi_1^{G_1}  = \sum_\s P^{\,\s }\,h( \s   )~,
\eea
we finally get
\be
W(C ; G_1 , G_2 )= \sum_{n_1=0}^\infty\,
{1\over {g_1 g_2 }}\,[N]^\g  \sum_{\sigma \sigma'}
P^\s C^{\s'}\,\d\left( \O^{1-2G_2} \Pi_1^{G_2} 
 ( h ( \sigma) \cdot h ( \sigma^{\prime} ) )     \right).
\ee
We defined 
 $\g=n_1+n_2-2(n_1G_1+n_2G_2)=(2-2G)n_1+n_S (1-2G_2)$, 
where we used  $n_2 = n_1 + n_S $.

\section{On the role of quantum characters in $q$-deformed 2d YM }\label{sec5}

In this paper we have used quantum $U_q(SU(N)$ characters rather than
 classical $SU(N)$
 characters. For the computations
in \cite{aosv} it seemed enough to consider classical $SU(N)$ characters.
 So one can ask: does one need to compute with
quantum characters, or do the classical ones suffice? In this section we
 argue that quantum characters are needed in the generic
situation; in fact, they are extremely natural and they provide the
 simplest solution to
 the problem of crossings and gluing along open lines. Our arguments
are consistent with \cite{aosv}, where the dimensions appearing in the
 partition function \eq{qpartf} 
were quantum dimensions but the characters 
associated with boundaries and Wilson loops were classical $SU(N)$ 
characters. In particular, this paper did not consider crossings 
on the surface, and gluing constructions involved closed curves only.
 In the absence of crossing 
points, both the classical and the quantum 
characters lead to a topological invariant theory. It is a well-known
 fact from Chern-Simons
 theory that one can do without $R$-matrices or other
quantum group structure as long as one considers simple Wilson loops
 -- for example, toric ones, 
whose expectation value  follows from 
surgery. In the 2d Yang-Mills case, the basic gluing formula along
 circles is \eq{glue},
which is valid both for classical and quantum characters, and
 ensures topological invariance of
 the gluing construction along circles. More 
precisely, the need for quantum characters in $q$YM can be seen:\\
\\
1) in the presence of Wilson loops with non-trivial crossings;\\
2) when gluing along open lines.\\
\\
The original definition of $q$YM is well-known \cite{br,boulatov} and it involves 
quantum characters. In the following subsections we collect several
 arguments that show the need for quantum characters.

\subsection{Consistency of Wilson loops}

One of the basic consistency conditions to be imposed on a Wilson loop is that, if the
 charge of the particle is zero, the expectation value
of the Wilson loop should be that of the unit operator; in other words, it should give
 back the partition function of the theory. In our case, if 
$W_R(G;C)$ is the Wilson loop operator in representation $R$ around the curve $C$ on the Riemann 
surface of genus $G$, consistency requires
\be 
W_{R=\r}(G;C) =\bra 1\ket=Z_{\sm{$q$YM}}(\S_G)
\ee
where $\r$ is the Weyl vector labeling the trivial representation. Thus, we should reproduce:
\be\label{partf}
W_{\r}(G;C)=\sum_S(\mbox{dim}_q\,S)^{2-2g}q^{-{1\over2}A\,C_2(S)}~.
\ee
We will check whether quantum dimensions and classical characters are consistent with this for a
 Wilson loop with crossings.

\begin{figure}
\begin{center}
\includegraphics
[height=2.2in,width=3in]
{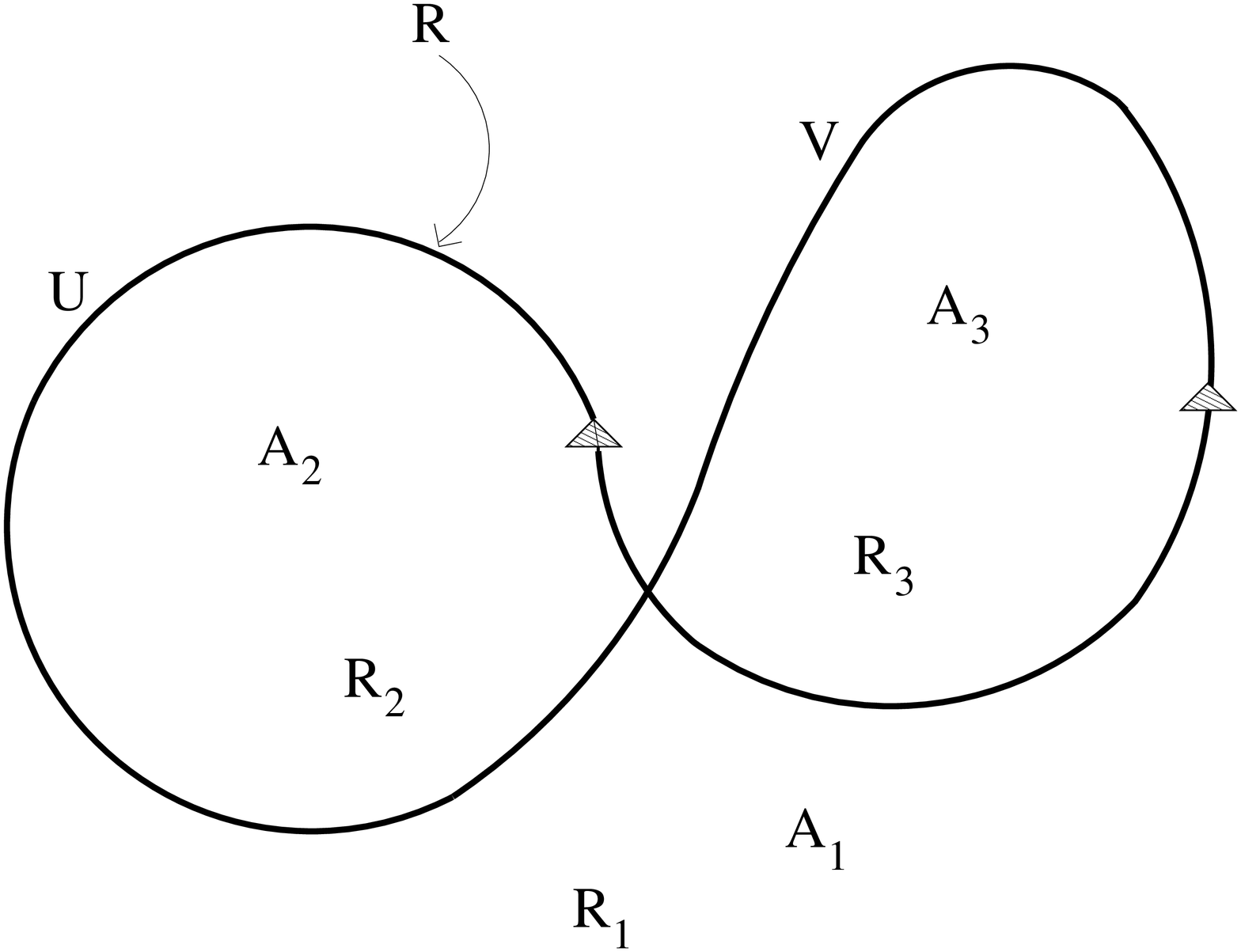}
\caption{\small A Wilson loop with a crossing.}
\label{fig8}
\end{center}
\end{figure}

Consider the expectation value of the Wilson loop $W_R(G;C)$ in Figure \ref{fig8}. In this case we have $A=A_1+A_2+A_3$, where $A_1$ is the area of the outer face, which  genus $G$. We get:
\bea
W_R(G;C)&=&\sum_{R_1R_2R_3}(\mbox{dim}_q(R_1))^{1-2g}\,\mbox{dim}_q(R_2)\,\mbox{dim}_q(R_3)\,
q^{-{1\over2}\, (A_1C_2(R_1)
+A_2C_2(R_2)+A_3C_2(R_3))}\times\nn
&\times&\int\dd U\dd V\,\chi_{R_1}(U^{-1}V^{-1})\,\chi_{R_2}(U)\,\chi_{R_3}(V)\chi_R(UV^{-1})
\eea
where, since we are dealing with classical characters, $\dd U$ is the Haar measure. Let us compute this in the trivial case: $R=\r$. 
We can compute the integrals using the character formula
\be\label{gluec}
\int\dd U\,\chi_{R_2}(U)\chi_{R_3}(U^{-1}V)=\d_{R_2R_3}{\chi_{R_2}(V)\over\mbox{dim}(R_2)}
\ee
We get
\be
\sum_S{(\mbox{dim}_q(S))^{3-2g}\over\mbox{dim}(S)}\,q^{-{1\over2}AC_2(S)}
\ee
which disagrees with \eq{partf}. The reason that the dimensions do not come out right
 is that we were forced to use the formula \eq{gluec}. We conclude that this procedure is 
not consistent. On the other hand, the same computation can be carried out with quantum
 characters, and in that case we do get the quantum dimension in \eq{gluec}.

\subsection{Gauge invariance of Wilson loops}

There is a short proof of gauge invariance for the Wilson loops and boundary elements we have discussed in previous sections. Let $U\in \mbox{Fun}_q(SU(N))$ (for more details on this see appendices \ref{appB} and \ref{appD}), and consider the ad-action of Fun$_q(SU(N))$ on itself:
\be
\mbox{ad}:U\mapsto h\,US(h)~,
\ee
where we are considering Fun$_q(SU(N))$ as a Hopf algebra with antipode $S$ \cite{frt}. It is easy to see that the quantum trace 
\be
\Tr\,(u\,U)
\ee
is left invariant under this action (for the definition of the $u$-element, see appendix \ref{appB}). We get:
\bea
\Tr\,(u\,hUS(h))&=&\Tr\,(S^2(h)uUS(h))=(S^2(h))_{ij}(uU)_{jk}(S(h))_{ki}\nn
&=&(S^2(h))_{ij}(S(h))_{ki}(uU)_{jk}=S(h_{ki}(S(h))_{ij})(uU)_{jk}\nn
&=&\Tr\,(u\,U)~,
\eea
where we used $\epsilon(h)=1$, and the fact that $u$ satisfies
\be
u\,x=S^2(x)\,u
\ee
for any $x\in\mbox{Fun}_q(SU(N))$. Thus, gauge invariance in Fun$_q(SU(N))$ is ensured provided we include the $u$-element.

We have proven that the triple
\be
\left({\over}\mbox{Migdal gluing~,~quantum dimensions~,~classical characters}\right)
\ee
is inconsistent in the generic case. To get a consistent theory, we need to modify one of the above. 
If instead of quantum dimensions we use classical dimensions, we of course get back the usual 2d Yang-Mills.
 If we want the dimensions to be quantum, we either need quantum characters, or a modification of the gluing
 rules. The possibility to have quantum characters has been discussed at length in this paper, and it has been
 shown to be consistent in \cite{br}. In particular, the theory is gauge invariant and independent of the
 triangulation. We do not exclude that there might be a complicated modification of the gluing rules that
 would allow to keep quantum dimensions and classical characters even in the presence of crossings. 

Additional features of the quantum characters are the following. The natural
 expansion of the quantum dimensions is in terms of quantum characters, which
 are most easily expressed in terms of a Hecke algebra, as we have shown. This
 gives a natural deformation of the symmetric group description of 
covering maps 
of the Riemann surface. Also in the case with boundaries the use of
 quantum characters 
was essential for this. Finally, $q$-deformed 2d Yang-Mills 
computes invariants of 
knots in Seifert manifolds \cite{SdHNPB,Sebaslps}. This is also expected 
from open-closed string 
duality in the A-model with branes. This relation will however only 
work if on the $q$YM
 side we deform the gauge symmetry as well so as to get quantum characters,
 since only that
 will give the quantum 6j-symbols that appear in the Reshetikhin-Turaev
 invariant
 relevant for knots in Chern-Simons \cite{Sebaslps}.

\section{Discussion and outlook } \label{discussion}

We have shown that the chiral large $N$ expansion 
(  note that the $q$-number $[N]$ appears as the natural expansion parameter )
for $q$-deformed Yang-Mills
can be described by Hecke algebras. The full large $N$ expansion is expected 
to be given by a coupled  product of chiral and anti-chiral contributions. 
We expect that techniques of this paper can be extended  to give a precise 
description of this non-chiral expansion in terms of Hecke algebras.

The string interpretation of $q$-deformed 2d Yang-Mills 
on $\Sigma_G$ has been 
developed in \cite{bp,aosv}.  The leading order terms 
in the expansion, obtained by setting the $\Omega$  factors
to $1$,  were shown to  compute Gromov-Witten invariants of a 
Calabi-Yau space $X$ which is a direct sum of line bundles 
$L_{p} \oplus L_{2g-2-p} $   fibered  over 
$\Sigma_G$. The sub-leading terms, due to the $\Omega $ factors
 were intepreted in terms 
D-brane insertions at $2G-2$ points. This picture develops 
  the Gross-Taylor interpretation  (at $q=1$)  of the $\Omega $ factors
  in terms of fixed points on the Riemann surface \cite{gti,gtii}. 
  An alternative interpretation  of the $\Omega$ factors underlies the 
  topological string theory developed in \cite{cmri,cmrii} for $q=1$.
 The latter  topological string is different from the
 standard one. It has been  labelled a balanced topological string  
 and has been  observed to be an example of a  general class
 of balanced topological  field theories naturally related 
 to Euler characters  of moduli spaces \cite{dijkmoore}. 
 It  integrates  over the moduli space of holomorphic maps
  the Euler class of the tangent bundle to that moduli space. 
 
The concrete connection between  Euler characters and the large $N$
 expansion of two dimensional Yang-Mills is manifest 
 when  one  expands the $\Omega $ factors and recognizes the binomial 
 coefficients as Euler characters of configuration spaces of 
 points on the Riemann surface  $ \Sigma_G $ \cite{cmri}. 
  Our treatment of the $\Omega $ factors
 in the $q$-deformed case, which has expressed it in terms of 
central elements of the Hecke algebra, naturally lends itself 
to this interpretation.  Euler characters of configuration spaces 
continue to appear in the expansion for the same reasons as at $q=1$. 
This suggests that a closed topological string interpretation exists 
for the large $N$ expansion of $q$-deformed two-dimensional 
Yang-Mills in terms of a balanced topological string.
 The simplest proposal along these lines 
is that the balanced topological string with target space  
$X$ would give a closed string interpretation  for the all orders expansion of 
 $q$-deformed two-dimensional Yang-Mills.
 The relation of such a picture to the D-brane insertions 
of \cite{aosv} would involve an interesting incarnation 
of open-string/closed-string duality. Developing these 
relations requires a clearer understanding of the 
coupling between holomorphic and anti-holomorphic sectors
in the context of the balanced topological string. The connection 
between the Gross-Taylor expansion and the Gromov-Witten invariants 
appearing in \cite{bp,aosv} has also been discussed in 
\cite{ccgpssi,ccgpssii}.

Given the rather simple Hecke $q$-deformation we have uncovered, 
of the sums over symmetric group delta functions related to the
 classical Hurwitz counting of branched covers,
 it is also natural to speculate that
there is an intrinsically two-dimensional picture 
which would account for the Hecke delta functions, without appealing to 
the Calabi-Yau $X$.  
One possiblity is that we have $q$-deformed Riemann surfaces
 and maps between 
such Riemann surfaces. In fact  $q$-deformed planes, known as Manin planes, 
 have been studied and holomorphy has been discussed ( see for example \cite{bdr}). 
One could construct Riemann surfaces which, in some sense, locally look like 
Manin planes, and consider holomorphic maps between them.
  As far as we are aware, 
such a theory of Hurwitz spaces for $q$-deformed Riemann surfaces 
has not yet been developed. 

While Hecke algebras are more familiar to mathematical physicsists
 as centralizers of quantum groups acting in tensor spaces, 
they have another pure mathematical origin ( see for example \cite{krieg}).   
$H_n(q)$  is an algebra of double cosets  $ B_n(F_q)
 \setminus GL_n ( F_q)/ B_n  (F_q) $. 
Here $F_q$ is the finite field with  
$q$ elements, where $q$  is a power of a prime $p$ ( If $q=p$ then $F_q$
 is just the field of residue classes modulo $p$ ).   $GL_n(F_q)$ is the group 
of $n \times n $ matrices with entries in $F_q$. $B_n(F_q)$ is the subgroup 
of the upper triangular matrices. This generalises the fact that 
$S_n$ appears from double cosets  $ B_n ( \mathbb{C} )
 \setminus GL_n ( \mathbb{C} ) / B_n ( \mathbb{C } ) $. 
Hence the deformation of  $ \mathbb{C} S_n$ to the Hecke algebra  $H_n (q) $
 corresponds to going from  $\mathbb{C}$    to $ F_q $. 
This suggests that, at least for $q$ equal to a power of a prime,
 our Hecke-$q$-deformed Hurwitz counting problem, 
 might be related to Riemann surfaces over $F_q $. It is interesting
 that, in this context, 
 fundamental groups can be defined and they still take the form 
\bea 
a_1\,b_1\,a_1^{-1}\,b_1^{-1}~ a_2\,b_2\,a_2^{-1}\,b_2^{-1}\, \cdots\,
 a_G\,b_G\,a_G^{-1}\,b_G^{-1}~ u_1 \cdots u_B =1 ~.
\eea 
 There are also results on the moduli spaces of branched
 covers in this set-up, 
 generalizing properties of classical Hurwitz space \cite{fulton69}. 
 An interesting direction for the future is to determine if there is 
 a relation  between Hecke algebras $H_n(q)$ and 
 these moduli spaces, and if such
 a relation provides the geometrical meaning for the $q$-deformed Hecke
 counting problems in (\ref{qcgt3}), (\ref{qcgt4}).

Classical and $q$-deformed 2d Yang-Mills are closely connected to Chern-Simons theory
on Seifert manifolds \cite{SdHjhep,Sebaslps,aosv,BW,BT}. On the other hand,
some of the formulas in this paper,  such as  \eq{tensorprod} ,  are suggestive of some connection of the
chiral large $N$ expansion of $q$-deformed 2d Yang-Mills and
orbifold conformal field theories \cite{dvvv} or Chern-Simons theory for finite
gauge groups \cite{dw,fq}. It is known that the Chung-Fukuma-Shapere three-dimensional topological 
field theory \cite{cfs} is the absolute value squared of the partition function of the Dijkgraaf-Witten 
theory. It seems very likely that the chiral expansion in terms of Hecke characters worked out in 
this paper can be formulated in the two-dimensional topological field theory framework of \cite{fhk,cfs}
with additional insertions coming from the branch points. It would be interesting to see in detail to what extent
the chiral $q$-deformed 2d Yang-Mills theory is related to the Dijkgraaf-Witten theory.
In view of the connection to Chern-Simons theory, it will be interesting to 
explore the $q$-deformed chiral as well when $q$ approaches roots of unity. $q$-Schur Weyl duality 
 at roots  of unity has been discussed in \cite{martin,blm}.

\bigskip
\section*{Acknowledgments }

\addcontentsline{toc}{section}{Acknowledgements}

We thank Mina Aganagic,   Luca Griguolo,   Costis Papageorgakis, 
Gabriele Travaglini, and Ivan Todorov for useful discussions and
 correspondence. SR  is supported by a  PPARC Advanced Fellowship.
The research of SR and SdH  is  in part supported by the
 EC Marie Curie Research Training Network MRTN-CT-2004-512194. 
 The work of AT is supported
by DFG (Deutsche Forschungsgemeinschaft) within the ``Schwerpunktprogramm Stringtheorie 1096''.
We have used the package  NCAlgebra for some of our computations with 
the Hecke algebra.

%\vfill

\begin{appendix}

\section{Central elements }\label{centralelements}

\subsection{Centrality of $q$-deformed conjugation sum } 
 We want to show that 
\bea\label{cent}  
 \sum_s  q^{-l(s) } h(s) h(t) h(s^{-1}) 
\eea 
 is central in $H_q(n) $. Since $H_q(n) $ is generated
by $g_1, ... g_{n-1} $, it suffices to show that the above element 
 commutes with these generators. 
We will first  show it for $g_1$, and it will be clear the same proof 
can be repeated for $g_2$ etc.  

First recall how this works in the case $q=1$. 
We write 
\bea 
 \sum_s s_1 sts^{-1} &&=  \sum_s ( \tilde s ) t { \tilde s}^{-1}  s_1  \nn 
&& =    \sum_{\tilde s } { \tilde s } t { \tilde s }^{-1} s_1~, \nm
\eea 
where we defined $ \tilde s = s_1 s $. 
The cancellation only uses a pair of terms at a time. For a fixed 
$s$, 
\bea 
&& s_1 s t s^{-1} = {\tilde s } t {\tilde s}^{-1} s_1 \nn 
&& s_1 { \tilde s } t { \tilde s }^{-1} = s t s^{-1} s_1 ~,\nm 
\eea      
which means that 
\bea 
[ s_1, sts^{-1} ] + [ s_1 , { \tilde s } t { \tilde s }^{-1} ] = 0 ~.
\eea 
It turns out that 
{ \it  the same pairwise cancellation works for $q \ne 1 $. } 
  It is instructive to check it explicitly for 
 $n=3,4$. Below we give the 
 general argument.  

Suppose $s$ is of the form $s_1 u $, where $u$ is a word in the 
generators. Now recall that before applying the map $h$ to $s$ we must express 
it in reduced form. This means that if $s =s_1 u $, the leftmost term in $u$ 
is not $s_1$. 
The following can be derived easily 
\bea 
&& h(s) = g_1h(u)  \nn 
&&  l(s) = l(u) +1  \nn 
&& h(s^{-1} ) = h(u^{-1}) g_1~. \nm
\eea   
Then $ \tilde s = s_1 s = u $. Now we write the pair of elements 
from (\ref{cent}) for the fixed $s, \tilde  s $. 
\bea\label{twoterms}  
&& q^{-l(s) } h(s) h(t) h(s^{-1} ) = q^{-l(u)  -1 }  g_1 h(u) h(t) h(u^{-1} )
 g_1 \nn
&&  q^{-l(\tilde s) } h(\tilde s) h(t) h(\tilde s^{-1} ) = q^{-l(u)} h(u) h(t) h(u^{-1} ) ~.
\eea 
The commutator with the first term  is
\bea\label{fstcom} 
&& [ g_1 , q^{-l(s)} h(s) h(t) h(s^{-1} ) ]  \nn 
&& = q^{-l(u) } h(u)h(t)h(u^{-1})g_1 + q^{-l(u) -1 } (q-1) g_1 h(u)
h(t)h(u^{-1})g_1 \nn 
&& ~~ - q^{-l(u) } g_1 h(u) h(t) h(u^{-1})  - q^{-l(u)-1} (q-1) g_1 h(u)
 h(t) h(u^{-1}) g_1~.
\eea 
The commutator with the second term in (\ref{twoterms}) is 
\bea\label{secom} 
 [ g_1 , q^{-l(u)} h(u) h(t) h(u^{-1})] = q^{-l(u)} g_1 h(u) h(t) h(u^{-1}) 
                     - q^{-l(u)} h(u) h(t) h(u^{-1}) g_1 ~.
\eea   
Combining the terms in (\ref{fstcom}) and (\ref{secom}) 
we see that the terms proportional to a power of $q$ cancel 
between the two equations ( as they must for this to work at $q=1$ ). 
The terms containing a factor $q-1$ cancel within (\ref{fstcom}). 

This proves that the sum (\ref{cent}) commutes with $g_1$. 
It has been done by decomposing the sum over $S_n$ into a sum over 
left coset elements  by the subgroup  $S_2$ generated by $s_1$, 
and a sum over representatives in each coset. The 
vanishing of the commutator with $g_1$ works  within the sum over
representatives in each coset. To prove that it commutes with
  $g_2 \cdots  g_{n-1} $ 
we similarly decompose with respect to left cosets of $s_2, \cdots s_{n-1} $.
 Hence  (\ref{cent}) is central in $H_q (n)$. It follows that it's 
matrix representation in any irreducible representation must be 
diagonal. Using the matrices given in \cite{kingwyb}, 
  we have checked this explicitly up to $n=4$.

A special case of (\ref{cent}) is given by the choice 
$t=1$. 
Based on evidence described below, we conjecture that its character 
in an irreducible representation is    
\bea\label{sumconjct} 
\sum_s q^{-l(s) } { \chi_R \over d_R(1) }  ( h(s^{-1} )  h(s) ) 
={  g ~ d_R(1)  \over d_R(q) }  ~,
\eea 
with $d_R(q) $ and $g$ as given in (\ref{dgforms}). 
 Since the Hecke element in the  character is
 central ( after summation over $s$ ), it suffices 
 to calculate it on one state in the irrep.
 We have checked this for general completetly symmetric
 reps and completely antisymmetric reps, 
as well as for all representations up to $n=4$, using the explicit 
matrices given in \cite{kingwyb}. 
Another check of this formula is to multiply by 
$d_R (q) d_R(1) $ and sum over young diagrams $R$ with $n$ boxes. 
Using (\ref{qdelchar}),  the LHS becomes 
\bea 
 g  ~  \delta \left(\!\!{\over} \sum_s q^{-l(s)}\, h(s^{-1} ) h(s) \right).
\eea 
But from  \cite{gyoja}  $ \delta ( h(s^{-1} ) h(s) ) = q^{l(s) } $. 
Hence the LHS is equal to $ ( g\, n!  )$. 
On the RHS we have $ g \sum_R  ( d_R (1))^2 = (g\, n!) $ . 
This gives a  consistency check of (\ref{sumconjct})  for any $n$.

Using (\ref{sumconjct}) and (\ref{characterfusion})
\bea 
  \sum_s  q^{-l(s) }   { \chi_R \over d_R(1) }  (  h(s) h(t)   h(s^{-1})  )& =& \sum_s q^{-l(s) } { \chi_R \over d_R(1) }  (  h(s)  h(s^{-1}) h(t )    ) 
\nn 
& =& { g ~ d_R(1) \over d_R(q) }\,  { \chi_R \over d_R(1) }  ( h(t) ) \nn 
& =& { g \over d_R(q) }\,   { \chi_R ( h(t )) } ~.
\eea 
Hence 
\bea 
  \sum_{s,t}  q^{-l(t)-l(s) }&&\!\!\!\!\!\!\!\!\!\!\!\!\!\! { \chi_R \over d_R(1) } ( h(s) h(t)
 h(s^{-1 } ) h(t^{-1}) )=\nn
& =& \sum_{s,t} q^{-l(t) } q^{-l(s) } \,{ \chi_R \over d_R(1) }  (  h(s) 
 h(t )    h(s^{-1})  )
  { \chi_R \over d_R(1) } (h(t^{-1} ) ) \nn 
& =& { g  \over d_R(q) }
    \sum_{t}  q^{-l(t) } \,{ \chi_R  } (  h(t) )
                  { \chi_R \over d_R(1) } (   h(t^{-1}) ) \nn 
&= &{ g  \over d_R(q) d_R (1) }\, \sum_{t}    q^{-l(t) }\, { \chi_R  } (  h(t) )
                 \chi_R  (   h(t^{-1}) ) \nn 
&=&  { g   \over d_R(q) d_R(1) }\,  { g d_R (1) \over d_R(q) } 
=  \left( { g \over d_R(q) } \right)^2 ~.
\eea 
The last sum over characters was done by using orthogonality 
(\ref{orthogrel}). This shows the desired identity (\ref{qdefpirel})

\subsection{Centrality of $q$-deformed commutator sum } 

 We prove that the element 
\be\label{centconj}  
C  \equiv \sum_{s,t } q^{-l(s) -l(t) } h(s) h(t) h(s^{-1} ) h(t^{-1}) 
\ee 
of the Hecke algebra $H_n(q) $ is central. In the $q=1$ limit, this is
$ \sum_{s,t} sts^{-1} t^{-1}$,  a sum of commutators of all 
group elements. Hence $C$ is a $q$-deformed  sum of commutators. 
Since $H_n(q) $ is generated by $g_1 \ldots g_{n-1} $ it suffices to 
prove that $g_i C = C g_i $ for any $g_i$. We will start with $g_1$ 
and it will be clear how to generalize to the other generators. 

Given the centrality of the $q$-deformed conjugation sum 
(\ref{cent}) we can write 
\bea 
 \Delta_1   &\equiv& g_1 C - C g_1  \nn 
& =&   \sum_{s,t } q^{-l(s) -l(t) } h(s) h(t) h(s^{-1} )  g_1 h(t^{-1}) 
 -   \sum_{s,t } q^{-l(s) -l(t) } h(s) g_1 h(t) h(s^{-1} )   h(t^{-1})~. \nn 
\eea 
We want to prove $\Delta_1 = 0$. 
For $q =1$ this can be proved as follows. If we define 
$ t = \eht s_1 ,  s = \hs s_1 $, we can write 
\bea 
\sum_{s,t}  ss_1 ts^{-1}  t^{-1}  = 
\sum_{ \hs , \eht  } {\hs }  \eht    {\hs}^{-1}  s_1 { \eht }^{-1} 
\eea 
This shows that it is useful to think about the sums over 
$S_n$ in terms of the cosets $S_n/S_2$ where the $S_2$ is generated 
by $s_1$. Let us choose expressions for the elements of $S_n$ 
in terms of words of minimal length in $s_1 .. s_n$. 
Let $S_+$ be the setof words not ending with $s_1$ on the right, 
and $S_-$ the set of elements of the form $ { \hat s } s_1$. 
Clearly $\hat s$   does not end with $s_1$ :  if 
it did $s$ would not be in reduced form. Hence $\hat s \in S_+$. 
For such $ s = { \hat s } s_1$, it  is easy to see that 
\bea 
&& h(s) = h ( \hat s ) g_1 \nn 
&& l(s) = l(\hat s ) +1 \nn 
&& h(s^{-1} ) = g_1 h( {\hat s }^{-1} ) 
\eea 
 We can write $\Delta_1 $ as 
\bea 
&& \Delta_1 =
(~  \sum_{s \in S_+ } + \sum_{s= {\hat s}s_1 \in S_-  ~ ; ~ { \hat s } \in S_+  } )   ~ ( ~  \sum_{t \in S_+ } + \sum_{t= {\hat t}s_1
 \in S_- ~;~ { \hat t } \in S_+ }  ) 
~ h(s) h(t) h(s^{-1} ) g_1 h(t^{-1} ) q^{-l(s) -l(t) } \nn 
&& - (~  \sum_{s \in S_+ } + \sum_{s= {\hat s}s_1 \in S_- ~;~ 
 { \hat s } \in S_+ }  )  ~ ( ~  \sum_{t \in S_+ } + \sum_{t= {\hat t}s_1
 \in S_- ~;~ { \hat t } \in S_+ }  )
~ h(s)g_1  h(t) h(s^{-1} )  h(t^{-1} ) q^{-l(s) -l(t) }  \nn
&& = \sum_{s,t \in S_+ } q^{-l(s)-l(t)} h(s)h(t)h(s^{-1})g_1h(t^{-1}) 
 + \sum_{s,{\eht} \in S_+ } q^{-l(s)-l( \eht )-1} h(s)h( \eht )g_1 h(s^{-1})
                                  g_1^2h({\eht}^{-1}) \nn 
&&  + \sum_{\hs, t \in S_+ }  q^{-l(\hs )-l(t)-1 } h( \hs )g_1h(t)g_1
h( {\hs}^{-1} )g_1 h( {t}^{-1} ) 
 + \sum_{\hs , \eht \in S_+ } q^{-l( \hs ) - l(\eht ) -2 } 
   h ( \hs ) g_1 h ( \eht ) g_1^2 h ( {\hs}^{-1}  ) g_1^2 h ( {\eht}^{-1} ) 
\nn 
&& - \sum_{s,t \in S_+ } q^{-l(s)-l(t)} h(s)g_1h(t)h(s^{-1})h(t^{-1})  
- \sum_{s,\eht  \in S_+ } q^{-l(s)-l( \eht )-1 } h(s) g_1 h(t)g_1 h(s^{-1}) 
                                  g_1 h(t^{-1}) \nn 
&& - \sum_{\hs , t \in S_+ } q^{-l( \hs )-l( t ) -1 } h( \hs )g_1^2 h(t )
g_1h({\hs}^{-1})h( t^{-1} )   
- \sum_{ \hs, \eht \in S_+ } q^{-l( \hs )-l( \eht ) - 2 } h( \hs )g_1^2 h( \eht )
g_1^2 h({\hs}^{-1}) g_1 h( t^{-1} )~. \nonumber 
\eea 
This can be simplified by using $g_1^2 = (q-1)g_1 + q $. 
We get terms with powers $q^{-l(s)-l(t) } $ in the summand but without 
powers of $(q-1)$, terms proportional to $(q-1)$ 
and terms proportional to $(q-1)^2 $. 
The terms without powers of $(q-1)$  cancel pairwise among the $8$ terms. 
The other terms can be written out explicitly, and seen 
to cancel. This proves that $[g_1,C]=0 $. 
When checking for commutation with $g_i$, we organise the sums
over $S_n$ according to cosets of the $S_2$ subgroup generated by $s_i$. 
Then the same argument as above applies to show that 
any of the generating $g_i$ commute with $C$. Hence $C$ is central.

\subsection{The elements $D$ and $E$ of $H_n(q)$ } 
The equation (\ref{sumconjct}) also allows us to give an expression for 
$D$ defined in (\ref{Delement}). 
Let us write 
\bea 
&& E = \sum_s q^{-l(s) }  h(s^{-1} )  h(s) \nn
&&   = 1 + \sum^{'}_{s}   q^{-l(s) }  h(s^{-1} )  h(s) \nn 
&& \equiv 1 + E^{\prime} \nm
\eea
The primed sum extends over elements 
in $S_n$ excluding the identity.
 Then we can write 
\be\label{defE}  
{ \chi_R ( E ) \over d_R(1) } = { g d_R(1) \over d_R(q) } 
\ee 
Using that $E$ is central 
\bea 
\left(  { \chi_R ( E ) \over d_R(1) } \right)^{m}
 = { \chi_R ( E^m  ) \over d_R(1) }  =  \left( { g\, d_R(1) \over d_R(q) } \right)^m 
\eea 
Now let $m=-1$ to get 
\be 
\chi_R ( E^{-1} ) = {d_R(q) \over g } 
\ee 
Hence
\bea  
&&  D = {g E^{-1}  }  \nn
&&  =  { g } \sum_{k=0}^{\infty}  (-1)^k  (E^{\prime} )^k \nn
&& =  { g } \sum_{k=0}^{\infty}  (-1)^k \sum_{u_1, u_2 \ldots u_k}^{\prime} 
                q^{-l(u_1) -l(u_2) - \ldots- l (u_k ) } 
 h(u_1^{-1} )  h(u_1)    \cdots  h(u_k^{-1} )  h(u_k) 
\eea 
 
\section{Quantum dimensions}\label{appB}

The irreducible representations $R$ of $U_q (U(N))$ 
can be realized as subspaces of $V^{ \otimes n } $, where 
$V$ is the fundamental representation. The matrix elements of 
the fundamental representation are denoted by $U$, with entries $U_i^j$ (see appendix \ref{appD} for explicit expressions), and the algebra 
generated by the $U$'s is dual to $U_q (U(N))$ and is denoted 
by $\mbox{Fun}_q ( U(N))$. The commutation relations of the $U_{i}^{j}$ 
are given in terms of the $R$-matrix in the referece we denote as FRT 
\cite{frt} (we are using $U$ 
 for $T$ of this reference).  

We first derive formula \eq{truh}, used in section 2 to obtain a Hecke formula for the 
quantum dimensions. Thus we need to compute the trace $\tr_n(u\,h(m_T))$ that comes from the 
quantum character expression. The element $u$ is:
\be \label{uelement}
u = q^{2 \sum_{i=1}^N  \left( { N+1\over 2 } - i \right) E_{ii} }~.
\ee  
The $E_{ij}$ act on the fundamental representation 
in the usual way
\be 
E_{ij}\, v_k = \delta_{jk}\, v_i~.
\ee

Now we can use the FRT formula for the $R$-matrix to show that 
\be 
 ( \tr \otimes 1 ) ( u \otimes 1 )\, P R = q^{N}\,  { \bf 1 } ~,
\ee
and $ \tr ( u )  = { q^{N} - q^{-N} \over q - q^{-1} } $. 
This means that 
\be 
( \tr \otimes \tr ) ( u \otimes u )\, P R 
= q^{N}\,{ q^{N} - q^{-N} \over q - q^{-1} }~.
\ee 
Going back to the Hecke algebra conventions using (\ref{grel}) 
($ q \rightarrow \sqrt{q }  $  ), we get 
\bea 
( \tr \otimes 1 ) ( u \otimes 1 )\,  g_1 &=& q^{N+1 \over 2 }\, \bf 1  \nn 
( \tr \otimes \tr  ) ( u \otimes u )\,  g_1 &=& q^{N+1 \over 2 }\, [N]~.
\eea 
More generally, tensor products of traces act on $u h(m_T)$ as
\bea\label{genfrmtr}  
 ( \tr \otimes \tr\otimes \ldots \otimes  \tr_{i} )
 ( u \otimes u \otimes \cdots \otimes  u ) ( g_1 g_2 \cdots g_{i-1}) 
   = q^ { (i-1) { N+1 \over 2 } }\, [N ]~.
\eea 
%We have used $ [N] =  { q^{N/2} - q^{-N/2} \over q^{1/2} - q^{-1/2}  } $. 

We now need to find out how to built $h(m_T)$ out of the $g_i$'s.
Consider a conjugacy class in $S_n$, denoted by $T$,  made of permutations 
which have $K_i$ cycles of length $i$. 
When expressed in terms of the 
generators $s_i$, the minimal length permutation in this conjugacy class, 
denoted by $m_T$, has length  $ \sum_{i} ( i-1 )K_i $.
  The minimal permutations are
given in terms of words of the form $g_{i} g_{i+1} ... g_{i +j } $, such 
as the one appearing in (\ref{genfrmtr}). For such minimal words, we can use 
(\ref{genfrmtr}) to obtain 
\bea\label{mintr}  
\tr_n ( u\, h ( m_T )  ) \equiv \tr^{\otimes n } \left(u^{\otimes n } h ( m_T )\right)
= q^{ { N+1 \over 2 } \sum_i (i-1)K_i    }\, [N]^{ \sum_i  K_i}  
= q^{   { N+1 \over 2 }\, l(m_T )}\, [N]^{ \sum_i  K_i}~.
\eea 
This is the formula \eq{truh} used in the derivation of the $q$-dimension formula in section 2. 

We now show explicitly how formula \eq{qcharexp} works in some examples, and that it leads to a 
$q$-dimension formula in terms of central elements \eq{qdimcent}.
%We consider the case $n=3$ explicitly, to illustrate the use of 
%the above traces and the steps leading to the $q$-dimension formula
%in terms of central elements (\ref{qdimcent}).  
For $q$-traces in $ V^{\otimes 3} $, i.e traces 
with $ u^{\otimes 3 } $ inserted, we have 
\bea 
\tr_q \,( 1 )   &=& [N]^3 \nn 
\tr_q \, ( g_1 )  &=& [N]^2\,  q^{N+1 \over 2 } \nn 
\tr_q\, ( g_1 g_2  ) &=& [N]\,  q^{N+1 }~,
\eea 
therefore
\bea \label{trg1g2g1}
  \tr_q ( g_2 g_1 g_2 )  &=&  ( q-1)\, \tr_q ( g_2 g_1 ) + q\, \tr_q ( g_1 )\nn
& =&  ( q-1)\, q^{N+1}\, [N]  + q\,q^{N+1\over 2 }\,[N]^2 ~.
\eea 
Now (\ref{qcharexp}) gives for the $q$-dimension 
\bea 
\mbox{dim}_q ( R ) &=& { 1 \over g } { d_R(q) \over d_R(1) }
    \left([N]^3 \chi_R ( 1 ) + 
 2q^{-1}\, q^{N+1 \over 2 }\, [N]^2\,   \chi_R ( g_1 ) + q^{-3}\, \chi_R ( g_1g_2g_1)\,  \tr_q ( g_2g_1g_2)\right. \nn
&+& q^{-2}\, \chi_R ( g_1g_2) \, q^{N+1}\, [N ] + q^{-2}\,  \chi_R ( g_2g_1)\,
 q^{N+1} [N ]\left.{\over}\!\! \right).
\eea
Filling in the above, we finally find 
\bea
\mbox{dim}_q ( R ) &=& { 1 \over g } { d_R (q) \over d_R(1) }
  \left( [N]^3\,  \chi_R ( 1 ) +  q^{N-1 \over 2 }\, [N]^2 
\chi_R  ( g_1 + g_2 + q^{-1} g_1g_2g_1 ) \right. \nn 
&+&\left.  q^{N-1 }\,[N]\, \chi_R ( g_1g_2 + g_2g_1 +q^{-1} (q-1) g_1 g_2 g_1 ){\over}\!\!\right)  \nn 
& =& { 1 \over g }\,   { d_R(q )\over d_R(1) }\, 
 \left([N]^3\, \chi_R ( 1 ) + q^{N-1 \over 2 }\,
 [N]^2\, \chi_R ( C_{T( 2,1)} ) + q^{N-1 }\, [N]\, \chi_R ( C_{T(3)}) \right).
\eea
The final expression contains central elements $C_T$
associated to conjugacy classes of $S_n$.  
There is the trivial conjugacy class containing 
the identity element,  for which $C_T (q) =1 $. 
There is $ C_{(2,1)} ( q ) = g_1 + g_2 + q^{-1} g_1 g_2 g_1 $, 
for the conjugacy class corresponding to a single transposition. 
Finally there is
 $ C_{(3)} ( q ) = g_1g_2 + g_2g_1 + { (q-1)\over q } g_1g_2g_1   $. 
It is easy to check that these elements commute with $g_1,g_2$. 
The above central elements and their generalizations are described 
in \cite{kac,dipjam}. They  approach the correct 
classical limit of a sum of permutations in the appropriate 
conjugacy class.

Using the Hecke characters given in 
\cite{kingwyb,ram}, we have checked that the above is consistent with the 
standard formula for the $q$-dimension as a product over the 
 cells of the Young diagram:
\be\label{qdimstandard}
\mbox{dim}_q\,R=\prod_{1\leq i\bra j\leq N}{q^{(\l_i-\l_j+j-i)/2}-q^{-(\l_i-\l_j+j-i)/2}\over q^{(j-i)/2}-q^{-(j-i)/2}}~,
\ee
where $\l_1,\ldots,\l_N$ are the lengths of the rows of the Young tableau, and, for $SU(N)$, $\l_N=0$. For $n=2$,
 an easy check, gives for the symmetric representation $\tableau{2}$ :
\be 
{ [N][N+1] \over [2] } 
\ee 
and for the antisymmetric representation $\tableau{1 1}$ :
\be 
{ [N][N-1] \over [2 ] }
\ee  
We have also obtained by the above manipulations, explicit formulae 
for central elements for $n=4$ which agree with those given 
in \cite{francis99}.  We have also checked that our formula the quantum dimensions \eq{qdimcent} agrees with the standard formula \eq{qdimstandard} for all representations up to $n=4$.

\section{Projectors}\label{appC}

Below we give explicit checks that (\ref{projform}) indeed 
defines projectors. We do this for $n=3$ and $n=4$, that is for the Hecke algebras of $H_3$ and $H_4$, and outline the method of \cite{gyoja} for the general case.

\subsection{$H_3$}

We work out the projector for a general representation of $H_3$. It contains $3!=6$ independent terms, corresponding to the six elements of 
$H_3$. Using \eq{trg1g2g1}, we get
\bea
P_R&=&{1\over c_R}\left(\chi_R(1) + {1\over q}\, \chi_R (g_1 )\,(g_1 + g_2) +\left({q-1\over q^3}\,\chi_R(g_1g_2) + {1\over q^2}\, \chi_R(g_1)\right) g_1g_2g_1+\right.\nn
&&\left. +{1\over q^2}\, \chi_R(g_1g_2)\, (g_1g_2+g_2 g_1)  \right)~,
\eea
where the term  $g_1g_2g_1$ corresponds to the $(13)$ permutation. Notice that in $S_n$, $s_1s_2s_1$ is in the same conjugacy class as $s_1$. Indeed, in the classical case where $q=1$ the first term in \eq{trg1g2g1} is absent and $\mbox{tr}(g_1g_2g_1)=\mbox{tr}(g_1)$. In the quantum case, $g_1g_2g_1$ has contributions from both $\chi(g_1g_2)$ and $\chi(g_1)$. This implies that it contributes to two different class elements.

Using the expressions for the characters in \cite{kingwyb}, we get for the three $H_3$ representations:
\bea
P_{\tableau{3}}&=&{1\over c_{\tableau{3}}}\left(1+g_1+g_2+g_1g_2+g_2g_1+g_1g_2g_1\right)\nn
P_{\tableau{2 1}}&=& {1\over c_{\tableau{2 1}}}\left(2 + {q-1\over q}\,(g_1 +g_2) - {1\over q}\,(g_1g_2+ g_2 g_1)\right)\nn
P_{\tableau{1 1 1}}&=&{1\over c_{\tableau{1 1 1}}}\left(1-{1\over q}\,(g_1+g_2) +{1\over q^2}\,(g_1g_2+g_2g_1) -{1\over q^3}\,g_1g_2g_1\right)
\eea
We have checked by explicit computation that they satisfy the projection equation \eq{proj}
provided
\bea\label{c3}
c_{\tableau{2 1}}&=&{q^2+q+1\over q}\nn
c_{\tableau{3}}&=&(q+1)(q^2+q+1)\nn
c_{\tableau{1 1 1}}&=&{(q+1)(q^2+q+1)\over q^3}~.
\eea
This agrees exactly with the values given in \cite{gyoja}, equation \eq{c} below.

\subsection{$H_4$}

For $H_4$, the projector contains $4!=24$ independent terms. The projector is:
\bea
c_RP_R&=&a+b(g_1+g_2+g_3)+c(g_1g_2+g_2g_3+g_2g_1+g_3g_2)+dg_1g_3\nn
&+&f(g_1g_2g_3+g_1g_3g_2+g_2g_1g_3+g_3g_2g_1) +h(g_1g_2g_1+g_2g_3g_2)\nn
&+&k(g_1g_2g_1g_3+g_1g_2g_3g_2+g_1g_3g_2g_1+g_2g_3g_2g_1) +lg_2g_1g_3g_2 \nn
&+&m(g_1g_2g_1g_3g_2+g_2g_1g_3g_2g_1) +ng_1g_2g_3g_2g_1 +pg_2g_1g_3g_2g_1g_3~.
\eea
The coefficients $a,b,c,d,f,h,k,l,m,n,p$ depend on the representation. 
They are characters of $q^{-l(\sigma)} \chi ( h( \sigma^{-1} ) ) $ 
which can be simplified, using cyclicity and the Hecke relations, to 
\bea
a&=&\chi(1)~,~b=q^{-1}\chi(g_1)~,~c=q^{-2}\chi(g_1g_2)\nn
d&=&q^{-2}\chi(g_1g_3)~,~f=q^{-3}\chi(g_1g_2g_3)\nn
h&=&q^{-3}[(q-1)\chi(g_1g_2)+q\chi(g_1)]\nn
k&=&q^{-4}[(q-1)\chi(g_1g_2g_3)+q\chi(g_1g_2)]\nn
l&=&q^{-4}[(q-1)\chi(g_1g_2g_3)+q\chi(g_1g_3)]\nn
m&=&q^{-5}[(q^2-q+1)\chi(g_1g_2g_3)+q(q-1)\chi(g_2g_3)]\nn
n&=&q^{-5}[(q-1)^2\chi(g_1g_2g_3)+2q(q-1)\chi(g_1g_2)+q^2\chi(g_1)]\nn
p&=&q^{-6}[(q-1)(q^2+1)\chi(g_1g_2g_3)+q(q-1)^2\chi(g_1g_2)+q^2\chi(g_1g_3)]~.
\eea
Again, the mixing between different terms comes from using formulas like \eq{trg1g2g1} and is related to the contribution of a single term to different central elements. In the limit $q=1$, each of the $a,d,\ldots,p$ depend on a single character, the one corresponding to the conjugacy class of the element that $a,d,\ldots,p$ multiply in the projector.

Using computer algebra, we have checked that the above are projectors for the five $n=4$ representations, provided
\bea\label{c4}
c_{\tableau{4}}&=&(q+1)(q^2+q+1)(q^3+q^2+q+1)\nn
c_{\tableau{3 1}}&=&{(q+1)(q^3+q^2+q+1)\over q}\nn
c_{\tableau{2 2}}&=&{(q+1)^2(q^2+q+1)\over q^2}\nn
c_{\tableau{2 1 1}}&=&{(q+1)(q^3+q^2+q+1)\over q^3}\nn
c_{\tableau{1 1 1 1}}&=&{(q+1)(q^2+q+1)(q^3+q^2+q+1)\over q^6}~.
\eea

\subsection{The construction for $H_n$}

In the general case, the projector contains $n!$ elements. Gyoja has given a formula for the coefficients\footnote{We corrected a typo in the formula in \cite{gyoja}.} $c_R$:
\be\label{c}
c_R={\prod_{i=1}^m(q-1)(q^2-1)\ldots(q^{\l_i+m-i}-1)\over\prod_{1\leq i\bra j\leq m}(q^{\l_i+m-i}-q^{\l_j+m-j})}\, q^{{1\over6}m(m-1)(m-2)}(q-1)^{-n}~.
\ee
From \eq{c3} and \eq{c4} we easily see that the coefficients satisfy
\be
c_R(q^{-1})=c_{R^{\tiny{T}}}(q)~,
\ee
where $R^{\tiny{T}}$ is the representation with transposed Young tableau. This is indeed a general property of the projectors \eq{c} \cite{op}.

It is easy to see that in the classical limit,
\be
c_R(q=1)={\prod_{i=1}^m l_i!\over\prod_{1\leq i\bra j\leq m}(l_i-l_j)}
\ee
where $l_i=\l_i+m-i$. This is the coefficient of the Young symmetrizer, and is given by the hook formula. Also the quantum coefficients \eq{c} can be expressed in terms of a hook formula.

For high $n$, it is tedious to check idempotency of the projector. Also, it relies on having explicit formulas for the characters of the Hecke algebra. Gyoja \cite{gyoja} has given a construction to compute projectors in general without recourse to characters. In this construction, to associate a projector to a particular representation $R$, we first associate a projector to every state of the representation. Every state is represented by a standard tableau $T$. A standard tableau is a tableau where the entries (numbered with elements from $\{1,\ldots,n\}$) are increasing across each row and down each column. The number of states in a given representation is $d_R(1)$. Thus, $P_R$ will be a sum of $d_R(1)$ primitive projectors, which we call $E_T$, where $T$ is the standard tableau they correspond to. The construction proceeds by defining two special tableaux, $T_+$ and $T_-$. These are the tableaux where the entries of the tableaux are numbered from 1 to $n$ successively across the first row (column), then the second, third, etc. $I_+$ and $I_-$ are the subgroups of $S_n$ that 
preserve the rows (columns) of $T_+$ ($T_-$). We associate to them parabolic subgroups $W_{\pm}$ of $S_n$  and define
\bea
e_+&=&\sum_{w\in W_+}h(w)\nn
e_-&=&\sum_{w\in W_-}(-q)^{-l(w)}h(w)~.
\eea
The primitive projector (up to normalization) associated to $T$ is then
\be
E(T)=h_-e_-h_-^{-1}h_+e_+h_+^{-1}
\ee
where $h_+=h_+(T)$ and $h_-=h_-(T)$ are the elements of the Hecke algebra corresponding to the permutation that transforms $T_+$ (resp. $T_-$) to the standard tableau $T$. Gyoja showed that the $E$'s are idempotents. The projector is then the sum of the orthogonal primitive idempotents:
\be\label{projg}
P_R={1\over c_R}\,\sum_{i=1}^{d_R(1)}E(T_i)~,
\ee
where $c_R$ was given before\footnote{Gyoja showed that the primitive idempotents are orthogonal using a certain ordering. In order for \eq{projg} to be a projector, they must be orthogonal independently of the ordering. This can be done defining new primitive idempotents in terms of the old ones, see Theorem 4.5 in \cite{gyoja}. For $n$ up to 4, however, we found that the primitive projectors are automatically orthogonal.}. We checked the previously constructed projectors for $n$ up to 4 using this construction. The first non-trivial case for $n=3$ is the representation $\tableau{2 1}$. There are two standard tableaux: $T_+=[\{1,2\}\{3\}]$ and $T_-=[\{1,3\}\{2\}]$. The permutation relating both is $(23)$, which is $h((23))=g_2$. In this case the parabolic subgroups are $W_+=W_-=\{1,s_1\}$, and 
\bea
e_+&=&1+g_1\nn
e_-&=&1-{1\over q}\,g_1
\eea
We further have $h_+(T_+)=1$, $h_-(T_+)=g_2$, therefore
\be
E(T_+)=1+g_1+{q-1\over q}\,g_2-{1\over q}\,g_1g_2+{q-1\over q}\,g_2g_1-{1\over q}\,g_1g_2g_1~.
\ee
For $E(T_-)$, $h_+=g_2$ and $h_-=1$, so
\be
E(T_-)=1-{1\over q}\,g_1-g_2g_1+{1\over q}\,g_1g_2g_1
\ee
The primitive idempotents are automatically orthogonal. We get
\be
P_{\tableau{2 1}}={q\over q^2+q+1}\left(E(T_+)+E(T_-)\right)={q\over q^2+q+1}\left(2+{q-1\over q}\,(g_1+g_2)-{1\over q}\,(g_1g_2+g_2g_1)\right)~,
\ee
in agreement with the formula obtained earlier.

As another example, we do $n=4$ for the representation $\tableau{3 1}$. There are three standard tableaux: $T_+=[\{1,2,3\},\{4\}]$, $T_-=[\{1,3,4\},\{2\}]$, and $T_3=[\{1,2,4\},\{3\}]$. We have $I_+=\{1,s_1,s_2\}$ and $I_-=\{1,s_1\}$, so
\bea
e_+&=&1+g_1+g_2+g_1g_2+g_2g_1+g_1g_2g_1\nn
e_-&=&1-{1\over q}\,g_1~.
\eea
In this case $h_+(T_+)=1$, $h_-(T_+)=g_3g_2$. Thus:
\be
E(T_+)=g_3\,g_2\,e_-(T)\,g_2^{-1}\,g_3^{-1}\,e_+(T)~,
\ee
which we worked out with the help of computer algebra. In the same way we have $h_-(T_-)=1$, $h_+(T_-)=g_2g_3$, so
\be
E(T_-)=e_-(T)\,g_2\,g_3\,e_+(T)\,g_3^{-1}\,g_2^{-1}~.
\ee
For $T_3$, $h_-(T_3)=g_2$, $h_+(T_3)=g_3$, hence
\be
E(T_3)=g_2\,e_-(T)\,g_2^{-1}\,g_3\,e_+(T)\,g_3^{-1}~.
\ee
The projector is the sum of the three, with the appropriate coefficient, and it agrees with the one computed directly. Notice that the primitive idempotents were automatically orthogonal in this case as well.

\section{$q$-Schur-Weyl duality and $q$-characters }\label{appD}

 In this appendix, we explain concretely the relation 
 between quantum characters of the $q$-deformed $SU(N)$ and 
 the symmetric group, in the special case of $SU(2) $.
   We will use the quantum group conventions of 
 \cite{nomura} and \cite{reshknots}.

 We will use the formulae for matrix elements 
 of spin-one representations from \cite{nomura} 
 in terms of spin-half representations and show that they 
 are consistent with expressing the characters in spin-one 
 in terms of the characters of spin half, using the Hecke algebra 
 generators, or $R$-matrices. For the $R$-matrix we will use the 
 notation of \cite{reshknots}. 

\subsection{$U_q(su(2))$ conventions}

We first summarize some of the formulas of \cite{nomura,reshknots} that we will use later.
The $U_q(su(2))$ algebra and coproduct are \cite{nomura}:
\bea 
 He - eH &=& 2e \nm\\ 
 Hf -fH &=& -2f \nm \\ 
  ef -fe   &=& {   q^{H/2} - q^{-H/2 }  \over   q^{1/2} - q^{-1/2} } \nn
 \Delta ( e ) &=& e \otimes q^{H/4} + q^{-H/4} \otimes e~.
\eea
For later convenience, we note that the map to the notation of \cite{reshknots}  is 
\bea 
q &\rightarrow& q \nn 
e &\rightarrow&  X_{+} \nn 
f  &\rightarrow&  X_-  \nn 
H &\rightarrow & H
\eea  
The universal $R$-matrix in this basis is \cite{reshknots}
\be 
R =  q^{ { { H \otimes H } \over 4 } }
  \sum_{n=0}^{\infty }   { ( 1 - q^{-1} )^n \over [n]! } \, 
 ( q^{H/4}X_+ )^n  \otimes  ( q^{-H/4} X_- )^n 
\ee 
where $[n]$ is as in \eq{qnumber}.
%$ [n] =  { ( q^{n/2}  - q^{-n/2} ) \over q^{1/2} - q^{-1/2} }  $. 
Together with the action of the generators on spin-half states,
\bea 
e \,|\half,-\half\ket &=& |\half, \half\ket \nn  
f \,|\half, \half\ket &=& |\half  ,-\half \ket  \nn  
H \,|\half, \pm  \half\ket &=&  \pm  |\half, \half \ket~,
\eea
this determines the $R$-matrix as follows:
\bea 
R_{\half,\half}^{\half , \half} &=&R_{-\half,-\half}^{-\half,-\half}= q^{1/4} \nn 
R_{\half,-\half}^{\half,-\half} &=&R_{-\half,\half}^{-\half,\half}=  q^{-1/4} \nn
R_{-\half,\half}^{\half,-\half} &=& q^{-1/4}\, ( q^{1/2} - q^{-1/2} ) ~.
\eea 

\subsection{Schur-Weyl duality in spin-one}

 As in the classical case, the $q$-characters in higher representations can be written 
 in terms of $q$-characters of lower representations. 
 Consider for concreteness the case of spin-one, which 
 is contained in the tensor product of two spin-half representations $V$.   
 There is a projector \eq{N=2proj} acting on $ V \otimes V $ that leads to the symmetric representation. In the classical case 
 it is just $ { 1 \over 2 } ( 1 + P ) $, where $P$ is the permutation 
of the two tensor factors. In the quantum case  $ P $
 does not commute with the coproduct, but 
 $ P R \equiv \check{R} $ does:
\be
\Delta \check{R} = \check{R} \Delta 
\ee
When $\check{R}$ acts on the tensor 
product of two spin half irreps, it  satisfies a 
relation of the form 
\be\label{krhrel} 
{\check{R}}^2 =  q^{-1/4} ( q^{1/2} - q^{-1/2} ) \, \check{R} +  q^{-1/2 }  
\ee 
A rescaling $g = q^{3/4} \check R $  can be done to map to the 
standard form of the Hecke algebra used in the main text. 
A matrix element of some element $h$ of $U_q ( su(2)) $   
 in the spin $1 $  representation can be written
in terms  of a product of spin half reps by using the Clebsch-Gordan 
coefficients.  
Consider now the following matrix element in the spin-one representation:
\be\label{spinonemat}  
\bra  j=1, n  | h | j=1, m\ket  
= d_{j=1;m}^{n } ( h ) =\bra h ,  d^{ { n }}_{j=1 ;  m } \ket
\ee 
$d_{j;m}^n$ is the representation matrix in representation $j$ with indices $n,m$, and $d_{j;m}^m$ its trace.
In the last equation we have expressed the fact that the 
matrix elements can be viewed as living in the dual space
 $U_q ( su(2)) $, denoted by $\mbox{Fun}_q ( SU(2)) $.  
For more details on this duality see for example 
\cite{majid,cosc9807,br}. 

We now express this in terms of matrix 
elements of the fundamental representation. They generate 
 $\mbox{Fun}_q ( SU(2) ) $, the deformed algebra of functions on $SU(2)$.   
Using the Clebsch-Gordan coefficients, we can rewrite the above as follows:
\bea\label{clebsh}  
\bra j=1, n  | h | j=1, m\ket &=& \sum_{m_1,m_2; n_1, n_2 }
 C_{n_1 n_2}^{n}\, C^{m_1 m_2}_{m}\,\times \nn 
&\times&  
%\hskip.3in 
\bra j=\half,n_1 |  \otimes 
 \bra j=\half , n_2 |  ( h_1 \otimes h_2 )
  |j=\half, m_1\ket \otimes  |j=\half, m_2 \ket
\nn 
&=&  \sum_{m_1,m_2; n_1, n_2 }
C_{n_1 n_2}^{n} \,C^{m_1 m_2}_{m} \,d_{j=\half; m_1}^{n_1} ( h_1 )  \,
d_{j=\half; m_2}^{n_2} ( h_2 ) \nn 
&=&  \sum_{m_1,m_2; n_1, n_2 }
C_{n_1 n_2}^{n}\, C^{m_1 m_2}_{m} \,
\bra h,  d_{j=\half; m_1}^{n_1} \,d_{j=\half; m_2}^{n_2} \ket ~.
\eea  
In the first equality, the co-product
 $ \Delta ( h ) = h_1 \otimes h_2 $ gives the action of $h$ 
 on the tensor product $ V \otimes V$. 
In the last equality, we used the fact that the dual pairing of a product of
two elements in $\mbox{Fun}_q ( SU(2)) $ is given by the co-product. 
Now we can sum over $m$ and use the identity between Clebsch-Gordan coefficients and projectors (see for example
 \cite{leclair})   
\be \label{CCP}
\sum_{ m } C_{n_1 n_2}^{m}\, C^{m_1 m_2}_{m} 
= P_{n_1 n_2}^{m_1 m_2} \left( \half,\half ; 1 \right) 
\ee
The projector is a linear combination of the identity and the 
$ \check{R} $. For $j=1$, the projector is in the tensor product of 
two spin-half representations. It has to be a linear combination of 
$1$ and $ \check R$ since the Hecke algebra generates the 
centralizer of the quantum group action in the tensor product:
\be\label{PabR}
P \left(\half,\half ; 1 \right)  = a + b \,\check R  
\ee
and for the matrix elements we have
\be
d_{j=1 ; m}^{n}  
= \sum_{m_1 , m_2 ; n_1 , n_2  } 
 d_{j=\half ; m_1 }^{n_1}\,  d_{j=\half ; m_2 }^{n_2}\, C^{m_1 m_2}_m \,C^n_{n_1 n_2} ~.
\ee 
To compute the character, we want the trace of this equation. Using \eq{CCP}, and expanding the projector in terms of the $R$-matrix as in \eq{PabR}, we get:
\be 
\tr_1 d  = a\, (\tr_{\half} d)\, (\tr_{\half} d) + b\, \tr_{1} ( { \check R} 
 ( d_{\half } \otimes 1 ) ( 1 \otimes d_{\half} )    )~.
\ee 
which, written out in indices, reads:
\be\label{sw1}  
\sum_{m  } 
 d_{j=1; m }^m 
= \sum_{m_1 , m_2 ; n_1 , n_2 } 
    \left ( a\, \delta_{n_1}^{m_1} \delta_{n_2}^{m_2} + 
 b \,R_{n_1 n_2}^{m_2 m_1} \right ) 
 d_{\half; m_1}^{n_1}\, d_{\half ; m_2}^{n_2} ~.
\ee
We will show that the above equation can indeed be solved 
for  constants $a,b$. The left-hand side can be calculated to give:
\bea\label{lh}  
\sum_{m  } d_{j=1; m }^m  &=& x^2 + ( xy + \sqrt{uv} ) + y^2 \nn
& =& x^2 + y^2 + xy ( 1 + q ) -q   
\eea 
where we have  used equations (36-40) of \cite{nomura}
 ( recalling that $x,y,u,v$ are the matrix entries of $d$ in 
the fundamental representation). 
For the right-hand side of (\ref{sw1}) we get 
\be\label{rh}  
( a + b q^{1/4} ) ( x^2 + y^2 ) 
 + a xy + 2b uv q^{-1/4} + ( a + bq^{-1/4} ( \sqrt {q} - 1/\sqrt{q} ) ) yx 
\ee 
Using the relations 
\bea
yx &=& (1-q) + q xy \nn
uv &=& q^{1/2} ( xy -1 ) 
\eea 
we can rewrite (\ref{rh}) in terms of $ x^2, y^2, xy , 1 $.  
 Comparing with (\ref{lh}) and considering the coefficient of 
$ x^2 + y^2 $ we immediately see that 
\be 
a + b\, q^{1/4} = 1~.
\ee 
With this condition the  coefficient 
of $xy$ becomes $ ( q+1 ) $ as desired. 
Comparing coefficients of the constant term 
then determines 
\be\label{ab}
a =  { 1 \over 1 + q }  ~,~ 
b =  { q^{3/4} \over 1 + q } 
\ee 
Putting everything together, and going back to the notation used in the main text, we get:
\be\label{newexp}  
\tr_{1}\, U = { 1 \over 1 + q } \,\tr\, U\, \tr\, U +  { q^{3/4} \over 1 + q } \,
\tr \otimes \tr   \left( \check R {\over} ( U \otimes 1 ) ( 1 \otimes U )  \right)  
\ee 
which is $q$-Schur-Weyl duality \eq{qtensexp} for $n=1$.
By comparing  (\ref{krhrel}) and with the first
of (\ref{hecke}) we can see that we can 
define $ g = q^{3/4} {\check R }  $. 
Then the projector can be read from above 
\be 
P_{\tableau{2}} = { 1 \over { 1 + q}} ( 1 + g ) 
\ee 
and agrees with (\ref{N=2proj})and the general form (\ref{projform}). 

\subsection{Quantum characters in spin-one representation}

The quantum characters can be obtained from the above by including the $u$-element \eq{uelement} in the trace, which is basically $q^{-H}$. In fact, we will do a slightly more general computation of the trace with an insertion of $ q^{ A H }$. Thus, we  consider the matrix element 
in the spin one representation of 
$ h q^{A H} $ where $A $ is an arbitrary number 
and $h$ is an arbitrary element o $U_q ( su(2) )$ :
\bea\label{spinonemat1}  
\bra  j=1, n  | h \,q^{A H }  | j=1, m\ket = q^{ A m }  d^ {n }_{j=1;  m} ( h ) 
\eea 
As before, we now rewrite this in spin-half matrix coefficients
 using the Clebsch-Gordan coefficients:
\bea 
\bra j=1, n  | h q^{ AH }  | j=1, m\ket &=& \sum_{m_1,m_2; n_1,n_2}  C_{n_1 n_2}^{n} \,C^{m_1 m_2}_{m}\,\times \nn 
%&& \hskip.3in
&\times&\bra j=\half,n_1 |  \otimes 
 \bra j=\half , n_2 |  ( h_1 \otimes h_2 )( q^{AH} \otimes q^{AH} ) 
   |j=\half, m_1\ket \otimes  |j=\half, m_2 \ket \nn 
&=&  \sum_{m_1,m_2; n_1, n_2 }
C_{n_1 n_2}^{n}\, C^{m_1 m_2}_{m} \, d_{j=1/2; m_1}^{n_1} ( h_1 ) \,d_{j=\half; m_2}^{n_2} ( h_2 )\,  q^{Am_1 + Am_2}  \nn 
&=&  \sum_{m_1,m_2; n_1, n_2 } C_{n_1 n_2}^{n} \,C^{m_1 m_2}_{m}\,q^{A m_1 + Am_2 } \,
\bra  h, d_{j=\half; m_1}^{n_1} \,d_{j=\half; m_2}^{n_2}  \ket ~.\nonumber
\eea  
Again, we can sum over $m=n$ to take the trace (notice the presence of $q^{AH}$ so this gives the quantum trace) and use \eq{CCP} to get 
\be 
\sum_{m} q^{ A m }\, d_{j=1; m}^{m} 
= \sum_{m_1 , m_2 ; n_1 , n_2 } q^{A ( m_1 + m_2)}
  \left ( a\, \delta_{n_1}^{m_1} \,
\delta_{n_2}^{m_2} + b\, R_{n_1 n_2}^{m_2 m_1} \right ) 
 d_{\half; m_1}^{n_1}\, d_{\half ; m_2}^{n_2} ~.
\ee 
For comparison to the classical formulae, it is 
useful to rewrite as 
\be\label{Anewexp}  
\tr_{1}\, ( q^{AH} U )  = a\, \tr\, ( q^{AH} U ) \,\tr\,  ( q^{AH } U )  
 + b \,\tr\otimes\tr\left(q^{AH} { \check R }{\over} ( U \otimes 1 ) ( 1 \otimes U )  \right).
\ee  
In the $q \rightarrow  1 $ limit,  $ \check R $ goes to 
the permutation $ P  $ and the second term becomes   
$ { 1 \over 2 }  \tr ( U^2 ) $. 
 
We still need to compute the constants $a,b$ in this case. Writing out the traces using \cite{nomura}, we get:
\bea\label{traceids}
\tr_1\,(q^{AH}\,U)&=&q^Ax^2+q^{-A}y^2+1+(q^{1/2}+q^{-1/2})uv\nn
\tr\otimes\tr\,\left(q^{A\,H\otimes H}{\over}(U\otimes1)(1\otimes U)\right)&=&(\tr\,(q^{AH}\,U))^2=
q^Ax^2+q^{-A}y^2+2+(q^{1/2}+q^{-1/2})uv\nn
\tr\otimes\tr\,\left(q^{A\,H\otimes H}\,\cR{\over}(U\otimes1)(1\otimes U)\right)&=&q^{1/4}[q^Ax^2+q^{-A}y^2+1-q^{-1}+(q^{1/2}+q^{-1/2})uv]\nn
\eea
where $A$ denotes an arbitrary power. It is now easy to see that the values of $a,b$ \eq{ab} are still the same,
 independently of the value of $A$. 

%Additionally, the explicit computation \eq{traceids} gives the following relations:
%\be
%(\tr\, q^{AH}U)^2-q^{-1/4}\,\tr\otimes\tr\left(q^{A\,H\otimes H}\,
%\cR{\over}(U\otimes1)(1\otimes U)\right)=1+q^%{-1}~.
%\ee
%Using \eq{traceids} and translating to RTF conventions, we have
%\be
%(\tr \,U)^2-q^{-1}\,\tr\otimes\tr\left((U\otimes1)(1\otimes U){\over}\cR\right)=1+q^{-2}~,
%\ee
%which implies
From the explicit computation \eq{traceids} we also get the special 
$N=2$ relations 
\bea
\Tr_{\tableau{1 1}}U&=&1\nn
\Tr_{\tableau{2}}U&=&(\tr\, U)^2-1~.
\eea
%[There should be an easy way to see this from antisymmetrization of tensors.] 
%We now go back to the general case.
%This reflects the fact that for $N=2$, any antisymmetric tensor in two indices
%can be replaced by an $\e$-tensor, and this holds also in the quantum case.

\end{appendix}

\end{document}